%% file: main_arXiv.tex
\documentclass{article}

\usepackage[margin=1in]{geometry}

\usepackage{amssymb}
\usepackage{amsmath}
\usepackage{booktabs}
\usepackage{verbatim}
\usepackage{enumitem}
\usepackage{subcaption}
\usepackage[colorlinks,citecolor=blue,urlcolor=blue,filecolor=blue,backref=page]{hyperref}

\usepackage[round]{natbib}

\newcommand{\vect}[1]{\boldsymbol{\mathbf{#1}}}
\newcommand{\todo}[1]{\textcolor{red}{#1}}

\usepackage{tikz}
\usetikzlibrary{bayesnet}

\newcommand\blfootnote[1]{%
\begingroup
\renewcommand\thefootnote{}\footnote{#1}%
\addtocounter{footnote}{-1}%
\endgroup
}

\begin{document}

\title{Author Clustering and Topic Estimation for Short Texts}

\author{ Graham Tierney$^\star$, Christopher Bail$^\dagger$, Alexander Volfovsky$^\star$ }

\maketitle

\blfootnote{$^\star$Statistical Science and $^\dagger$Sociology, Duke University}

\begin{abstract}
    Analysis of short text, such as social media posts, is extremely difficult because of their inherent brevity. In addition to classifying topics of such posts, a common downstream task is grouping the authors of these documents for subsequent analyses. We propose a novel model that expands on the Latent Dirichlet Allocation by modeling strong dependence among the words in the same document, with user-level topic distributions. We also simultaneously cluster users, removing the need for post-hoc cluster estimation and improving topic estimation by shrinking noisy user-level topic distributions towards typical values. Our method performs as well as --- or better --- than traditional approaches, and we demonstrate its usefulness on a dataset of tweets from United States Senators, recovering both meaningful topics and clusters that reflect partisan ideology. We also develop a novel measure of echo chambers among these politicians by characterizing insularity of topics discussed by groups of Senators and provide uncertainty quantification.    
\end{abstract}

\section{INTRODUCTION}

The proliferation of micro-blogs and other short text on social media platforms, product reviews, and online advertisements has increased interest in models that can extract useful information from short documents. 
Standard text analysis tools require large numbers of long documents to estimate useful quantities. One of the most widely used models for text data is the Latent Dirichlet Allocation (LDA) topic model from \cite{Blei2003LDA}, which estimates topics by within-document word co-occurrences. If document word counts are extremely sparse --- as is the case with short text --- LDA and other topic models will suffer in topic estimation. Moreover, when texts themselves are short, researchers often want to jointly model both text and other features such as similarities among the authors of the texts. 

We motivate the development in this paper by considering a dataset of all tweets by United States Senators from the first five months of 2020 and use it to measure partisanship across topics expressed in the tweets. This range covers the second impeachment of President Trump through the early stages of the coronavirus pandemic. Tweeting has become an important communication tool for US politicians who issue positions, interact with constituents, and attempt to set the political agenda on a daily basis \citep{Jaidka2019brevity}. Just as the introduction of radio addresses in 1923 and televised debates in 1960 marked transition points in American political discourse, Twitter use has now become a focal point of US politics. Legislators respond to topics raised by their constituents on Twitter \citep{barbera2018leads}, and President Trump used Twitter to distract from negative news stories \citep{lewandowsky2020using}. Clustering users in this political context can reveal the presence or absence of partisan topics, which can measure ``echo chambers'' and ideological segregation \citep{Bakshy2015,Bail9216}. Many analyses of echo chambers on social media focus on account-level homophily, e.g. identifying that Twitter users typically follow accounts with political ideology similar to their own \citep{Moslehe2021tie_formation,Barbera2015birdsfeather_tweet_together}. This tie formation would affect exposure to different kinds of political messages only if politicians discuss distinct topics. As such, our model focuses on the content of politician messages. 
However, characterizing behavior and topics discussed at the tweet-level is still extremely challenging due to the brevity of the messages. For example, in our application, the post-term matrix, where the $ij$ entry is the count of appearances of word $j$ in post $i$, is 99.8\% zeros. 

In this paper we develop a model that directly targets short text while leveraging similarities among users to improve topic coherence. Specifically, we cluster U.S. Senators by Twitter activity, and discover meaningful topics and groupings of accounts. We collected a dataset of all tweets by sitting US Senators from January 1st, 2020 to May 31st, 2020, which, after pre-processing described in Section~\ref{sec:application}, contains 61,241 tweets and 2,644 unique words. Our model captures partisan differences in coverage of the coronavirus pandemic and economic stimulus. Moreover, our clustering of Senators based only on their Twitter activity maps onto distinct ideologies that can transcend party identification as well: Senators who tweet similarly have similar political beliefs. Those Senators who are grouped with out-partisans are frequently outliers within their own party, such as the moderate Republican Senators Susan Collins and Lisa Murkowski and the Democratic Socialist Senator Bernie Sanders. We also quantify insularity of topics discussed by the identified clusters by measuring how many tweets it would take each cluster to mention every topic. We find that the more liberal Democrats and nationally-focused Republicans have the broadest topic coverage on average, but individual Democrats are much less varied than individual Republicans in these categories, which means that individual Democrats tend to cover every topic faster. In contrast to two-stage procedures that first learn topics then cluster users, our estimated topics are more coherent and our clusters better match established theoretical measures of Senator ideology \citep{poole2017ideology}. 

The most common method researchers have used to overcome the brevity of messages on social media platforms is to merge all short documents by the same user (or user type) into a single, long document and estimate a topic model on those combined documents \citep{hong2010empirical,pennacchiotti2011machine,steinskog-etal-2017-twitter,barbera2018leads}. More recent theoretical work has estimated which tweets to merge from the data by borrowing methods from the information retrieval literature \citep{HAJJEM2017761}.  
Central to these methods is the assumption that no information is contained in the fact that certain words are {\it chosen} to belong to the same short document. A Twitter user facing significant character-limits choosing to write two words in conjunction should imply that their topics are more similar than the same two words used in distinct tweets.

Given the downsides of combining short texts into a single long text per user, \cite{zhao2011comparing} first introduced a single topic per-short-text LDA (called Twitter-LDA) to compare the content of New York Times articles to Twitter. Each user tweeted about a mixture of topics, and each tweet had only a single latent topic.\footnote{Incorporating authorship information into topic models is not new; the author-topic model of \cite{rosen2004author} associated each author in a corpus with a distribution over topics that was was identified by multi-author documents. Multi-author documents, however, are typically not found in social media.} They found that using a single topic per tweet --- rather than a mixture of topics --- produced more semantically coherent topic distributions. This is very intuitive: it is highly likely that only a single message can be conveyed in under 280 characters. The Twitter-LDA model proposed by \citet{zhao2011comparing} is nearly identical to the Dirichlet-Multinomial Mixture (DMM) model in \cite{Yin:2014:DMM:2623330.2623715}. In the latter model, each user produces a single document, which is mapped to a single topic distribution over words. This is equivalent to single-topic LDA where each document comes from a unique user. DMM methods traditionally ignore user information, and model a single corpus-level mixture over topics, rather than a user-level mixture. Both DMM and Twitter-LDA were derived as collapsed Gibbs samplers, making them impractical for even moderate sample sizes. To overcome the issue that each text is produced by a unique user in DMM, the recently proposed LapDMM, which also derives a collapsed variational algorithm, introduces a regularization term that makes semantically similar texts, identified by word-embedding methods discussed later, more likely to have the same topic \citep{li2019dirichlet}. This procedure requires knowing or learning this semantic distance {\em a priori} which again suffers from the same constraints that all short text modeling suffers from.

Other data augmentation approaches have recently combined learned topic information from longer documents with shorter documents in terms of keywords \citep{Sahami2006webkernel} or actual text \citep{Hu2009wikipediatransfer,jin2011transferring}. A significant limitation of these methods is that they require a corpus of long text that is {\em a priori} compatible with the short-text documents being studied. \cite{Yan2013biterm} proposed a biterm model (BTM) for short text where topic-level word co-occurrence relationships are learned for the whole corpus. Each document is decomposed into the set of all two-word pairs (biterms) that can be formed from the document's contents. Each biterm is modeled as being drawn from a single topic. The lack of a meaningful generative model makes the results less interpretable and the method cannot leverage information that the same user generated multiple documents. Moving beyond topic models altogether,
neural networks have been introduced to estimate word embeddings in a latent vector space, then classify short texts based on a distance metric between texts \citep{mikolov2013efficient,rangarajan-sridhar-2015-unsupervised-topic}. The word mover's distance \citep{Kusner2015wmd} is one popular method to compute a distance matrix for a corpus. 
More recent work has leveraged word embedding methods to augment topic models, either with embeddings learned from a large external corpus \citep{li2016gpudmm} or the corpus of short documents being modeled \citep{Shi:2018:STM:3178876.3186009}. These methods generally ignore user information (the distance metric does not account for authorship), and, more importantly, rely on pre-trained word embeddings from larger corpora, which may not reflect the specific vocabulary and semantic usage of words in the corpus in question. This reliance especially limits the ability of these methods to capture new or emerging topics. 

Thematically similar to LapDMM, clustering techniques have been introduced into the LDA universe to improve the performance of topic modeling.
\cite{Barnard:2003:MWP:944919.944965} proposed a mixture of LDA models to group documents with an associated image. Within a cluster, each document's topic distribution is a draw from a cluster-specific Dirichlet distribution. \cite{wallach2008structured} proposed a similar clustering model that placed a hierarchical prior on the cluster-specific Dirichlet parameters. 
\cite{xie2013integrating} expanded that level of clustering to model global topics shared across all documents and local topics used only within a cluster. Just as with LapDMM, the clustering here is defined at the document level and not at the user level.

\subsection{Our contribution}
To improve upon these methods for short text we combine a generalization of the short text LDA topic model with unsupervised clustering of authors of short documents. {\it This hierarchical model is able to share information at multiple levels leading to higher quality estimates of per-author topic distributions, per-cluster topic distribution centers, and author cluster assignments}. 
Our method fuses the clustering of {\bf both users and documents}. We leverage the dependencies of words in a single post and borrow strength across users to inform user-specific topic distributions. Our method outperforms standard tools on simulated data even when our model is misspecified, and provides unique insights in a real-world application. 
To facilitate computation we derive both a collapsed Gibbs sampler that directly targets the posterior as well as a variational approximation to that posterior. We note that prior work on short text has heavily relied on slow and unscalable Gibbs samplers \citep{zhao2011comparing,Yan2013biterm} and that the LapDMM variational approximation heavily relies on a one document per-user assumption. Our method can handle multiple short texts per user and generalizes the variational approximation derived for traditional LDA such that it can be naively parallelized. We show that this approximation improves, in terms of inference being comparable to that of the Gibbs sampler, as the scale of the data increases. This computational gain extends the model to otherwise intractable datasets.

The remainder of the paper is organized as follows: Section~\ref{sec:model} outlines the details of the proposed short text LDA with clustering (stLDA-C) model and shows its' relationship to previously proposed models. Section~\ref{sec:parameter_estimation} derives the collapsed Gibbs sampler and the variational approximation to the posterior. 
We apply the model in Section~\ref{sec:application} to the dataset of US Senator tweets, identifying distinct clusters of politicians and heterogeneity in topics by partisanship. We leverage the form of our model to define a novel metric of the echo-chamber effect on social media and discuss the potential heterogeneity of this effect.
Section~\ref{sec:modelvalidation} validates the proposed model and estimation procedures on simulated datasets, both when the model is correctly specified and when it is misspecified.

\section{MODEL\label{sec:model}}

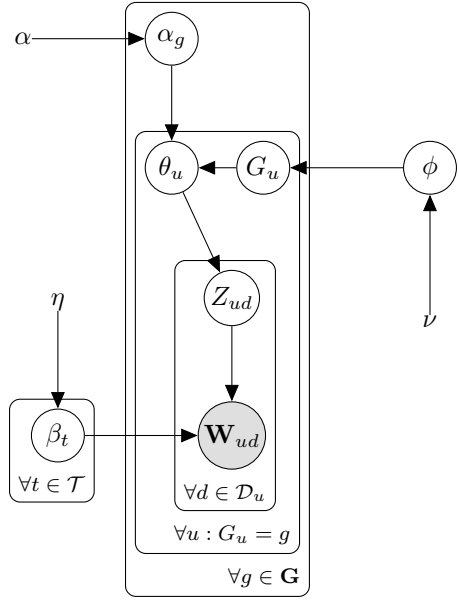
\begin{figure}[ht]
  \begin{center}
    \input{Floats/hstLDA_phi.tex}
  \end{center}
\caption{Plate diagram for the stLDA with clustering model. The diagram further illustrates the connection between the model proposed in this manuscript with previous models in the literature. Observed quantities are illustrated in gray and objects without a border around them are hyperparameters.}
  \label{fig:stLDA_graph}
  
\end{figure}
The general framework for latent Dirichlet allocation models specifies a probabilistic relationship between a collection of topics, a corpus of words, and the documents that are comprised of those words.
Traditional LDA (\citet{Blei2003LDA}, referenced as tLDA below) models each document as a mixture over topics: Each word has its own latent topic assignment, and documents have unique topic distributions. The tLDA model conceptualizes topics by considering words co-occurring in the same document, but when the documents are short those co-occurrences are extremely rare. This baseline approach further ignores the information about whether documents were produced by the same author, which is particularly common for short text from social media platforms. Our proposed model captures this user-level dependence and improves topic estimation by leveraging the dependence of the latent topic variables of words in the same tweet. 

Concretely, our method adds \emph{three} elements to the traditional LDA model. To model short texts: (1) we restrict each document to a single topic, rather than a mixture over topics. (2) We allow mixing of topics at the user-level, where observed author information allows us to learn more accurate topics by borrowing strength across documents authored by the same user. Lastly, (3) we incorporate unsupervised clustering to group users by similarity in their topic choices. This allows for information sharing between users with similar topic distributions and shrinkage of user-topic distributions with high uncertainty towards a common mean. We refer to our model as stLDA-C. 

Figure \ref{fig:stLDA_graph} displays the plate diagram of our hierarchical model. Prior distributions over cluster-level topics, user cluster assignments, and topic distributions over words are governed by fixed parameters $\alpha$, $\nu$, and $\eta$ respectively. $\alpha_g$ is the vector parameter of a Dirichlet distribution over topics choices for users in cluster $g$, identified by $G_u = g$. Each user-specific topic distribution $\theta_u$ is a draw from $\text{Dir}(\alpha_g)$. For each tweet $d$ by user $u$, its topic, $Z_{ud}$, is a single draw from $\theta_u$, and all words in tweet $ud$ are sampled from the topic distribution over words, $\beta_t$, where $Z_{ud} = t$. $\phi$ encodes the proportion of users in each group and forms a prior distribution for $G_u$. The generative process is described below: 

\begin{enumerate}[topsep=0pt,itemsep=-1ex,partopsep=1ex,parsep=1ex]
    \item For each cluster $g=1,\ldots,G$, sample $\alpha_g \overset{\text{iid}}{\sim} f_\alpha$, where $f_\alpha$ is the prior over $\alpha_g$
    \item For each topic $t=1,\ldots,T$, sample $\beta_t \overset{\text{iid}}{\sim} \text{Dir}(\eta)$
    \item Sample $\phi \sim \text{Dir}(\nu)$
    \item For each user $u$:
        \begin{enumerate}[topsep=0pt,itemsep=-1ex,partopsep=1ex,parsep=1ex]
            \item Sample $G_u \sim \text{Cat}(\phi)$: $P(G_u=g) = \phi_g$
            \item Sample $\theta_u \sim \text{Dir}(\alpha_{G_u})$
            \item For each document $d=1,\ldots,n_u$
                \begin{enumerate}%[topsep=0pt,itemsep=-1ex,partopsep=1ex,parsep=1ex]
                    \item Sample the topic $Z_{ud} \sim \text{Cat}(\theta_u)$ 
                    \item Sample words $\vect{W_{ud}} \sim \text{Multi}(n_{ud},\beta_{Z_{ud}})$
                \end{enumerate}
        \end{enumerate}
\end{enumerate}

\noindent Multi(n,p) is the multinomial distribution of size n and probability vector p. Cat(p) is the categorical distribution, a multinomial of size one. Dir($\zeta$) is the Dirichlet distribution with parameter $\zeta$. Below we discuss how our model generalizes aspects of other LDA models and specializes them to short text. 

\input{other_methods}

\section{PARAMETER ESTIMATION\label{sec:parameter_estimation}}

We develop two estimation procedures: first, an exact but (potentially) slow collapsed Gibbs sampler and, second, a faster but approximate variational inference method. 

\subsection{Collapsed Gibbs\label{sec:collapsed_gibbs}}

We estimate the latent topic parameters $Z_{ud}$ with collapsed Gibbs sampling. \cite{Griffiths2004ldagibbs} developed the collapsed Gibbs sampler for tLDA by analytically integrating out $\theta_d$ and $\beta_t$, so only samples of the latent topic variables were required. Below we derive the collapsed sampler for stLDA-C. We are interested in $P(Z_{ud} = t|\vect{Z_{-ud}},\vect{W},G_u=g,\alpha_{g})$, where $\vect{Z_{-ud}}$ denotes the topics of all tweets except for tweet $d$ by user $u$, $\vect{W}$ denotes the word counts for all tweets, and $G_u$ and $\alpha_{g}$ contain the cluster assignment and parameters for $u$. Conditional on $\theta_u$ and $\beta_t$, there is no dependence on topics or words in other tweets. Hence we can write $P(Z_{ud}=t|\theta_u,\beta_t,\vect{Z_{-ud}},\vect{W}) \propto P(\vect{W_{ud}}|Z_{ud}=t,\beta_t)P(Z_{ud}=t|\theta_u) \propto \prod_i \beta_{ti}^{W_{udi}} \theta_{ut}$, where the product is taken over all $i=1,\ldots,V$ unique words in the corpus. This density can be integrated over the posteriors for $\beta_t$ and $\theta_u$ given the words and topics of other tweets and the $\alpha_g$ hyper-parameters. With independent Dirichlet priors on all $\beta_t$ and hierarchical Dirichlet priors on $\theta_u$, the posteriors are also Dirichlet by conjugacy of the Dirchlet and Multinomial distributions. As such we have $\theta_u|\vect{Z_{-ud}} \sim \text{Dir}(\alpha_g + \vect{Z_{-d}^{(u)}})$, where $\vect{Z_{-d}^{(u)}}$ is the vector of topic counts in only user $u$'s tweets excluding tweet $d$ and $\beta_t|\vect{Z_{-ud}},\vect{W} \sim \text{Dir}(\eta + \vect{W_{-ud}^{(t)}})$ where $\eta$ is the parameter of the prior and $\vect{W_{-ud}^{(t)}}$ is the word counts of tweets with topic $t$ excluding tweet $ud$. Putting these together we get:
\begin{align*}
    p(&Z_{ud} = t|\vect{Z_{-ud}},\vect{W}) 
    = \int \theta_{ut}p(\theta_u|\vect{Z_{-ud}})\ d\theta_u  \int \prod_i \beta_{ti}^{W_{udi}} p(\beta_t|\vect{Z_{-ud}},\vect{W_{-ud}}) \ d\beta_t 
\end{align*}
These integrals are equivalent to the posterior predictive probabilities of observing $Z_{ud}=t$ and $\vect{W_{ud}}$ given the topics and words of other tweets. The integral over $\theta_u$ is the same as in \cite{Griffiths2004ldagibbs} and it simplifies to: $$\frac{\alpha_{gt} + (Z_{-d}^{(u)})_t}{\sum_j \alpha_{gj} + (Z_{-d}^{(u)})_j}$$ 
The second integral simplifies to: 
$$\frac{\Gamma\left(N\eta + \sum_i (W_{-ud}^{(t)})_i\right)}{\prod_i \Gamma\left(\eta + (W_{-ud}^{(t)})_i\right)} \frac{\prod_i \Gamma\left(\eta + (W_{-ud}^{(t)})_i + W_{udi}\right)}{\Gamma\left(N\eta + \sum_i (W_{-ud}^{(t)})_i + \sum_i W_{udi}\right)}$$
For more detailed proofs of this derivation see Appendix A. These two quantities are calculated for each topic, then normalized to sum to 1 across all topics. 

Corpus-level cluster proportions $\phi$ are estimated from conditionally conjugate prior distributions. Given the counts of cluster membership $\vect{G}$ and a $\text{Dir}(\nu)$ prior on $\phi$, $\phi|\vect{G} \sim \text{Dir}(\nu + \vect{G})$. $\phi$ provides a prior distribution for each $G_u$, the cluster indicator for user $u$. 

We note that $p(G_u=g|\phi,\vect{Z^{(u)}},\alpha_g) \propto P(\vect{Z^{(u)}}|\alpha_g,G_u=g)p(G_u=g)$. The first term is simply the Dirichlet-Multinomial probability of the given topic counts, and the second term is $\phi_g$. It is worth noting that the $\phi_g$ term effectively builds into the model a preference for fewer clusters. The posterior for $\phi_g$ will be roughly proportional to the number of other users in cluster $g$, which means that when sampling a new cluster for user $u$, the model will prefer to place $u$ into a larger cluster. This results in a property where some clusters will be empty if the data are better modeled by fewer clusters than the number the researcher has specified. We will demonstrate this property in our simulations below.

We sample $\alpha_g$ with a Metropolis update within the collapsed Gibbs sampler. Any prior distribution can be chosen for $\alpha_g$, but it is easiest to think of $\alpha_g$ as $\vect{m_g}c_g$ where $\vect{m_g}$ is a point on the simplex in $\mathbb{R}^T$ and $c_g>0$ is a concentration parameter. A Dirichlet prior on $\vect{m_g}$ can capture the prior expected value of $\theta_u$ and the prior on $c_g$ determines how concentrated the prior is around that expected value. An important consideration in estimation is if the concentration for group $g$ is small, very different user topic distributions have similar likelihoods in that group. Informative priors on this parameter can be useful to ensure that within-group topic variability is low. Alternatively, if a diffuse prior is used, the model may estimate that certain clusters have zero users in them while other clusters have highly variable topic distributions. This feature can be useful to determine if the data are more effectively described as coming from fewer clusters.

The most computationally demanding part of the MCMC is the collapsed sampler for $Z_{ud}$. The topic of each tweet depends on the topic assignment of all other tweets, even for tweets by other users, because those topics inform beliefs about $\beta_t$. Thus, this step does not parallelize across users or groups. Consequently, it is useful to compute many updates to the other parameters, $\alpha_g$, $\phi$, and $G_u$, after each iteration through $Z_{ud}$ to improve mixing.\footnote{Updating certain parameters more frequently does not change the stationary distribution of the Markov chain. It simply helps find the local optima of the re-sampled parameters given the topics.}

\subsection{Variational Inference\label{sec:variational_inference}}

Variational inference is commonly used to estimate LDA topic models \citep{Blei2003LDA}. The key difficulty in estimating the full posterior for stLDA-C, as with traditional LDA, is the link between $\beta_t$ and $\theta_u$. The dependence between these two parameters, clearly observed in the above Gibbs sampler, can lead to poor mixing and a computational bottleneck. The link is why the collapsed Gibbs sampler needs to update each post's topic sequentially rather than in parallel. In variational inference, the true posterior is approximated by a simpler distribution, typically with less dependence among the parameters. We follow the same approach and approximate the true posterior by minimizing the KL divergence between the variational distribution and the true posterior. 

We first describe the variational approximation for a single user, the analog of a single document in tLDA, then describe how that result generalizes. We wish to identify three latent, user-specific variables given higher-order parameters and the data: $p(g_u,\theta_u,\vect{Z}_u|\phi,\vect{\beta},\vect\alpha,\vect{W}_u)$. $g_u$ is user $u$'s cluster membership, $\theta_u$ is the user's topic choice probabilities, and $\vect{Z}_u$ contains the topics of the user's tweets. $\phi$ is the corpus-level cluster proportions, $\vect\beta$ is the $T \times V$ matrix of $T$ topic distributions over the $V$ words in the corpus, $\vect \alpha$ is the $G \times T$ matrix of $G$ cluster-specific, $T$-dimensional Dirichlet parameters. $\vect W_u$ is the $n_u \times V$ document-term matrix for user $u$. For clarity the $u$ subscript will be omitted for the remainder of this section. 

We approximate the distribution $p$ above with the variational distribution $q$: 
\[q(g,\theta,\vect{Z}|\lambda,\gamma,\xi) = q(g|\lambda)q(\theta|\gamma)\prod_{d=1}^{n}q(Z_d|\xi_d)\]
where $q(g|\lambda)$ is categorical over the different clusters with probabilities $\lambda$, $q(\theta|\gamma)$ is Dirichlet with parameter $\gamma$, and $q(Z_d|\xi_d)$ is categorical over the different topics with probabilities $\xi_d$. Note that these are also user-specific quantities. In this section, the parameters of the user-specific variational distributions are referred to as the variational parameters, and the parameters $\vect \alpha$, $\vect \beta$, and $\phi$ that are shared across users are referred to as model parameters. The variational inference approach (summarized by Figure~\ref{fig:vb_algo}) iterates between learning user specific parameters (in parallel over users) and updating the overall model parameters.

\begin{figure}[ht]
    \centering

\begin{enumerate}
    \item Initialize model parameters $\phi$, $\vect\alpha$, and $\vect \beta$.
    \item Initialize variational parameters $\lambda_u$, $\gamma_u$, and $\xi_u$.
    \item \textbf{repeat}
    \item \hspace{10pt} \textbf{for} each user $u$:
    \item \hspace{10pt} \hspace{10pt} Update variational parameters with Equations (\ref{vb:labmda}) - (\ref{vb:xi})
    \item \hspace{10pt} Update model parameters with equations (\ref{vb:phi}) - (\ref{vb:beta}) and Appendix B results. 
    \item \textbf{until} convergence
\end{enumerate}

    \caption{Full Variational Algorithm for stLDA-C}
    \label{fig:vb_algo}
\end{figure}

\subsubsection{Variational Parameter Estimation}

We want to set the variational parameters $\lambda$, $\gamma$, and $\xi$ to minimize the KL divergence from the variational distribution to the posterior. Using $D(q||p)$ to denote this quantity, the optimization problem can be expressed as: 
\[(\lambda^*,\gamma^*,\xi^*) = \underset{\lambda,\gamma,\xi}{\text{arg min}} D(q(g,\theta,\vect{Z}|\lambda,\gamma,\xi) || p(g,\theta,\vect{Z}|\phi,\vect{\beta},\vect\alpha,\vect{W}))\]

Detailed in Appendix B, the minimization can be computed with an iterative fix point method. The update rules are shown below. 
\begin{align}
    \lambda_g &\propto \phi_g \exp(E_q[\log p(\theta|g,\alpha_g)]) \label{vb:labmda}\\
    \gamma_t &= \sum_g \lambda_g \alpha_{gt} + \sum_d \xi_{dt} \label{vb:gamma}\\
    \xi_d &\propto \exp(E_q[\log \theta_t]) \prod_{j=1}^V \beta_{tj}^{W_{dj}} \label{vb:xi}
\end{align}

Next, we explain each equation and the interpretation of the terms where possible. The expectations of functions of $\theta$ are tractable and computed analytically.% in Appendix B. 
Equation ($\ref{vb:labmda}$) is the variational probability of cluster membership in cluster $g$. The update rule is interpretable as the prior $\phi_g$ times the likelihood of the user's topic distribution under cluster $g$, with the likelihood approximated by the exponential of the expected log-likelihood under the variational distribution. Equation (\ref{vb:gamma}) is the variational approximation to the user's topic distribution. The Dirichlet parameter for topic $t$ is the weighted average of each cluster's Dirichlet parameter ($\alpha_{gt}$), with weights given by the variational probability of cluster membership ($\lambda_g$), plus the total variational probability of posts by the user having topic $t$. This result is analogous to the variational update in \cite{Blei2003LDA} Equation (7) with the single model parameter $\alpha_i$ replaced with the cluster-weighted average. Equation ($\ref{vb:xi}$) is the variational probability that post $d$ by the user is from topic $t$. The update rule is interpretable as proportional to the prior $\theta_t$ times the likelihood of the post coming from topic $t$, with $\theta_t$ approximated by the variational distribution. This is again analogous to the variational update in \cite{Blei2003LDA} Equation (6), except because all words share the same topic, the post-likelihood term is multiplicative.  

A key result here is that because of the independence in the variational distribution, each user's update only depends on model and user-specific variational parameters. Thus, the operations can be computed in parallel if such computing resources are available. This is a significant computational improvement over the collapsed Gibbs sampler where the update of $\vect Z$, each post's topic, has to be computed in sequence. This, along with only needing a single value of the variational parameters, rather than a sample from the posterior, are what make this approximation method faster. 

\subsubsection{Model Parameter Estimation}

We can estimate the remaining model parameters with an empirical Bayes approach. The marginal likelihood $p(\vect W | \phi,\vect \alpha,\vect \beta)$ is intractable, but the variational posterior provides a lower-bound on the likelihood that we can use instead. Essentially, the general procedure is the same variational EM method as in variational estimation of traditional LDA. First, use the above results to find $\{\lambda^*,\gamma^*,\xi^*\}$ for each user. Second, compute the values of $\{\phi,\vect \beta, \vect{\alpha}\}$ to maximize the lower bound on the likelihood computed in the first step. Iterate between these steps until convergence. 

The derivations of these results are shown in Appendix B. Because these parameters are shared across users, we re-introduce the $u$ subscripts indicating user and $\mathcal{D}_u$ indicating the set of documents produced by user $u$. 
\begin{align}
    \phi_g &\propto \sum_u \lambda_{ug} \label{vb:phi}\\
    \beta_{tj} &\propto \sum_u \sum_{d\in\mathcal{D}_u} \xi_{udt} W_{udj} \label{vb:beta}
\end{align}

The update for $\vect \alpha$ is more complicated and does not have a clear intuition, so we omit the statement here. It is available in Appendix B and is computed via an efficient Newton-Raphson method very similar to traditional LDA. 

Equation ($\ref{vb:phi}$) is clearly interpretable as setting $\phi_g$ proportional to the total variational probability of users belonging to cluster $g$. Equation (\ref{vb:beta}) for $\beta_{tj}$ is similarly the total variational probability that word $j$ is assigned to topic $t$ across all users and documents. 

A particular concern with one-topic-per-word models is when attempting to model a new document, new words cannot be assigned a topic because $\beta_{tj'}=0$ where $j'$ is not in the training data. To address this issue, \cite{Blei2003LDA} extends the variational approximation to cover the topic distributions over words. They perform adaptive smoothing where the update rule is $\beta_{tj} \propto \eta_t + f(t,j)$, essentially adding a topic-specific constant $\eta_t$.  This additional approximation is not needed for the one-topic-per-document model proposed here. There is no need to assign a unique topic to word $j'$. The topic of the new document can be inferred from the words that appear both in the new document and in the training data. 

Note that when initializing model parameters, $\vect \alpha$ and $\vect \beta$, uniform starts are not advisable. If rows of $\vect \alpha$ are identical, then each user is equally likely to be placed in each cluster (the clusters are all identical), and the resulting $\lambda_u$ values will all be uniform. Similar results for $\xi_u$ will occur if $\vect \beta$ is initialized as uniform.

\subsection{Comparison\label{sec:vb_gibbs_comparison}}

In this section, we demonstrate the quality of the variational approximation to the true posterior, with additional comparisons to a very flexible topic model, the Biterm Topic Model (BTM, \citet{Yan2013biterm}), and the model commonly used in practice, what we refer to as combined LDA (cLDA). We find that the variational approximation improves as the size of the data increases, which is also when the computational advantage of the variational approximation is most useful. 

\begin{figure}[ht]
    \centering
    \caption{Gibbs and Variational Bayes (VB) Comparison}
    \begin{subfigure}{.40\textwidth}
        \includegraphics[width=\textwidth]{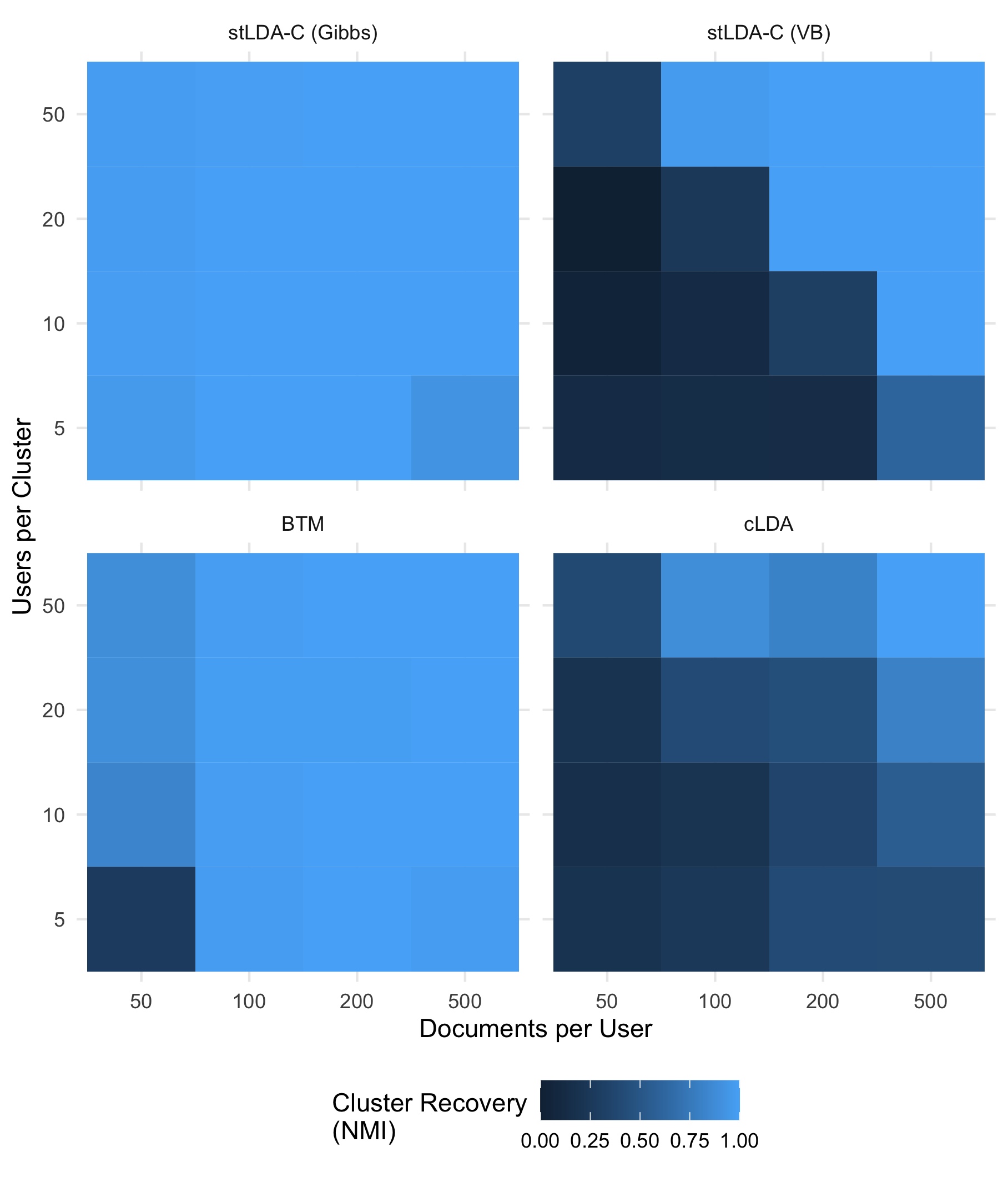}
        \caption{Cluster Recovery}
    \end{subfigure}%
    ~
    \begin{subfigure}{.40\textwidth}
        \includegraphics[width=\textwidth]{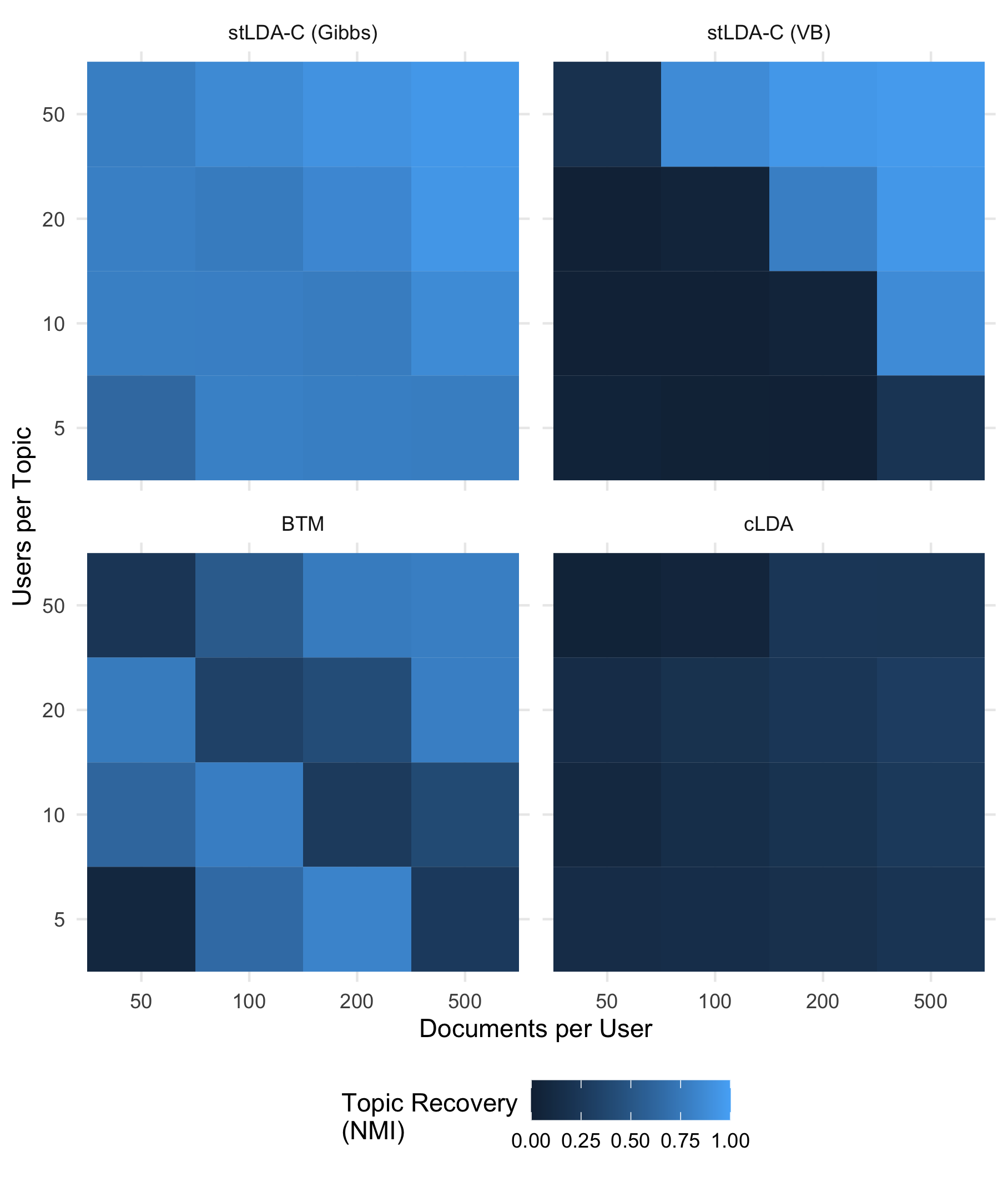}
        \caption{Topic Recovery}
    \end{subfigure}
    \label{fig:vb_gibs_comp}
    
        \footnotesize
        \flushleft
    Notes: Data was simulated from the stLDA-C model with three clusters, two that post about disjoint sets of five topics and one that covers all ten topics. The number of posts per user and number of users per cluster were varied for each simulation. Documents and users are classified into the cluster and topic with the highest posterior probability. Normalized Mutual Information is used to evaluate each classification. Gibbs estimation of stLDA-C fully recovers the truth. Variational estimation will recover the truth, matching the correct posterior, with at least 5,000 total documents. 
    Results are averaged over 7 replications for all but the largest setting, which was replicated only twice due to computation time for the Gibbs samplers. 
\end{figure}

Figure \ref{fig:vb_gibs_comp} plots cluster and topic recovery, showing that the variational approximation is quite close to the Gibbs estimation given sufficient data. We use stLDA-C as a generative model and simulate users from 3 clusters and 10 topics. Two of the clusters discuss disjoint sets of 5 topics, and the third cluster discusses all topics. This simulation is quite favorable to stLDA-C, with well-separated clusters and the single-topic per document assumption holds. We know the Gibbs sampler will quickly converge to the true posterior, which enables us to evaluate the quality of the variational approximation. The top-right panel in each sub-figure of Figure \ref{fig:vb_gibs_comp} shows that once the data are of sufficient size, the variationally-approximated and Gibbs-estimated posteriors match and recover the ground truth. Our recommendation is to only use the variational method when the amount of documents is prohibitively large, which is the exact scenario when the variational approximation is of high quality. BTM and cLDA are able to separate users by cluster fairly well, but still struggle with topic estimation.\footnote{Clustering for these models is performed after estimation and is described in detail in Section \ref{sec:modelvalidation}} More extensive model validation, with more comparison methods and simulations where stLDA-C is misspecified, are provided in Section \ref{sec:modelvalidation}. 

\subsection{Analyst set variables}

Two key parameters are chosen by the analyst, the number of topics, $T$, and the number of clusters $G$. In traditional topic models, $T$ is often chosen to be large, with the understanding that some topics will pick up on background noise and likely be uninterpretable. When each document covers multiple topics, this approach does not hinder inference because the high-probability topics are the ones that are most important and the less-interpretable ones tend to be lower probability across documents. In a topic-per-document model, this behavior is less desirable because the model may map an entire document to a non-interpretable or incoherent topic. We generally recommend choosing fewer topics than in traditional topic models, and potentially lowering the number of topics if many documents are mapped to non-interpretable topics. 

$G$ is also an important variable, and closely tied to $\nu$, the prior on the corpus-level proportions of each cluster. A large $\nu$ encodes prior belief that the number of users in each cluster will be similar, and small $\nu$ encodes prior belief that there will be one large cluster and several smaller clusters. Depending on the analyst's intention for downstream analysis using the clusters, this flexibility can be important. A key feature here as well is that the model can identify that fewer than the specified number of clusters can fit the data well. In simulations with very little heterogeneity across users, our model correctly identifies that users come from only one cluster and unique set of $\alpha_g$ parameters. As such, we recommend that analysts using this model think about setting $\nu$ specific to their prior beliefs and desired uses of cluster information, as well as potentially setting $G$ larger than desired to identify if clusters are emptied out. Various hold-out methods can be used to evaluate different choices of $T$ and $G$, where model parameters are learned on a training set of documents, then a measure of model fit is computed on a set of held-out documents. We demonstrate such methods in Sections \ref{sec:application} and \ref{sec:modelvalidation}. 

Identifiability in topic models, as in all latent factor models, is a concern. A rotation of the topics produces equivalent likelihoods, which could hinder inference and Gibbs sampling in particular. Literature on the identifiablity in topic models focuses on methods that use separable non-negative matrix factorization of the document-term matrix or some transformation of it \citep{anandkumar2013overcomplete,Huang2016}. One solution is to specify anchor words that appear in one and only one topic \citep{Donoho2004nmf}. When this assumption is not appropriate, other assumptions on higher-order moments of the corpus or topic distributions can suffice \citep{anandkumar2012svd_lda,anandkumar2013overcomplete,Huang2016}. 

Specifying anchor words requires strong domain knowledge, and they may be difficult to learn when the number of topics is large. These issues are less concerning in this single topic per (short) document model. One could specify anchor-documents with less domain knowledge because more context is provided; one only needs to identify a sets of documents that the analyst would like to separate into different topics. Per the above discussion, generally our method is better with fewer topics than in topic-per-word models, so this task might not be too onerous for the analyst. We do not rely on anchoring topics in our simulations or application, but the R code provided has functionality to pre-specify topics of a subset of documents. We use the Gibbs sampler only in Sections~\ref{sec:modelvalidation} and \ref{sec:vb_gibbs_comparison}, and we do not observe evidence of label switching in the posterior draws. Topic or cluster rotations would not affect the variational approximation used in Section~\ref{sec:application}.

\section{APPLICATION\label{sec:application}}

We collected a dataset of tweets from sitting U.S. Senators from January 1st, 2020 to May 31st, 2020. To form an analyzable dataset, some pre-processing steps are taken: all words are made lower case, then stop words, hyperlinks, and words used in fewer than 0.1\% of tweets are removed, and finally all documents with one or fewer words remaining are dropped. The final dataset includes 61,241 tweets covering 2,644 words. We set $T$, the number of topics, to 30 and $G$, the number of clusters, to 4 and proceed with estimation using the variational approximation derived in Section~\ref{sec:variational_inference}. Four clusters were chosen to allow for separation of the two parties and some within-party heterogeneity. We also compared topic coherence on a set of held-out tweets for models with 2, 4, 6, 8, and 10 clusters.\footnote{The hold-out set was created by randomly sampling 10\% of tweets from each Senator with at least 10 tweets, excluding only Senator Ben Sasse.} The model with 4 clusters had the best score. Topic coherence has been shown to correlate with human-judged semantic coherence and is a common evaluation metric for topic models \citep{mimno2011coherence}. Moreover, pure held-out likelihood measures are know to have \emph{negative} correlation with human-judged coherence \citep{Chang2009reading}. The four cluster model also had the best clustering NMI relative to party labels, reflecting the best separation of senators by partisanship. 

stLDA-C estimates meaningful topics: tweets from the same topic relate to similar legislative priorities and political events, with distinct partisan- and non-partisan topics. The clusters are also meaningful, mapping onto senator ideology measured by DW-NOMINATE scores. First, we will describe the topics estimated, then second discuss the senator clustering. 

\subsection{Estimated Topics}

\begin{figure}[ht]
    \centering
    \includegraphics[width=.75\textwidth]{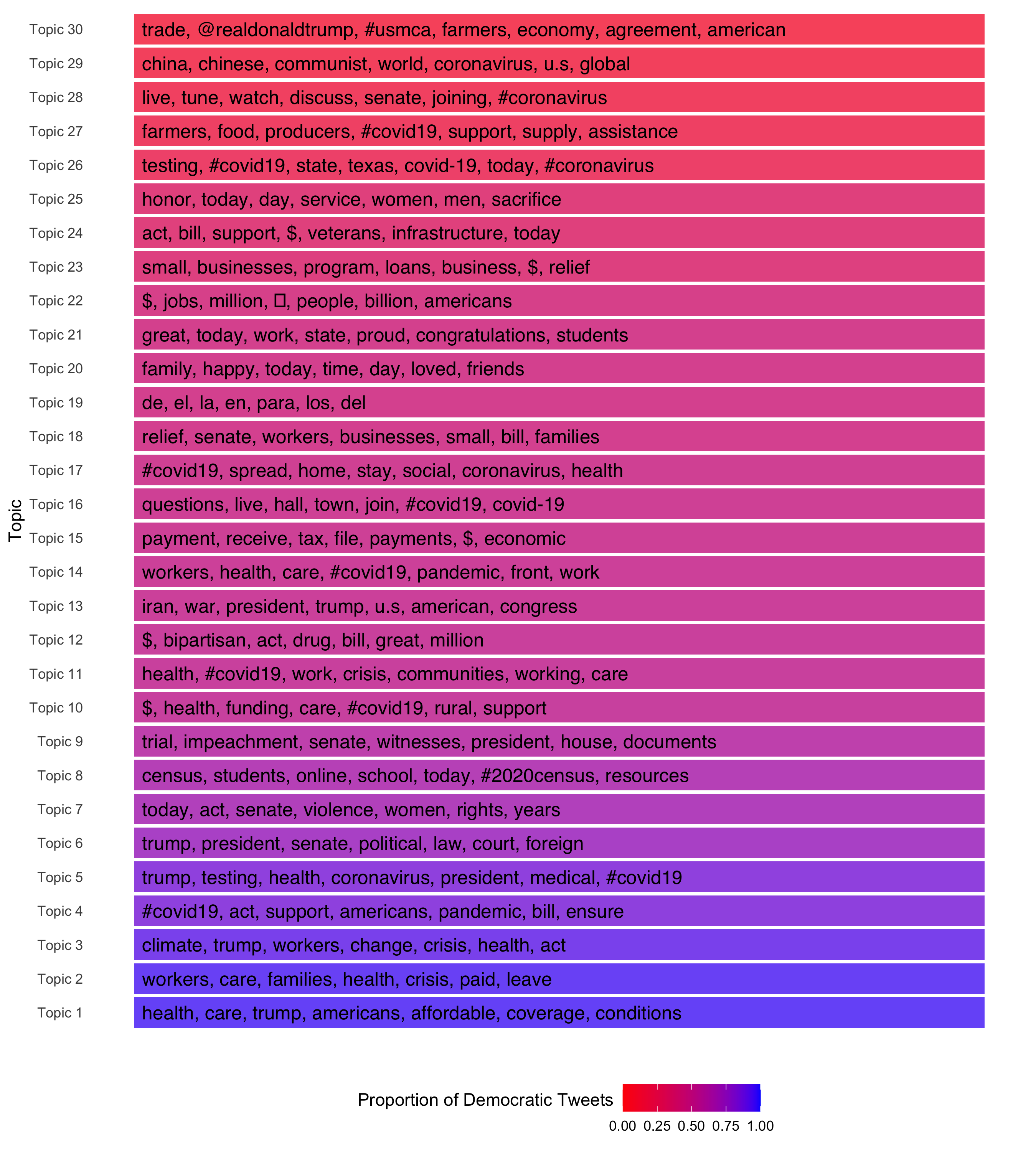}
    \caption{Top Topic Words and Topic Partisanship}
    \label{fig:app_topwords}
    
    \flushleft 
    \footnotesize
    Notes: Each topic is represented by its 7 most probable words. Topics are ordered and shaded by the percentage of tweets in that topic authored by Democratic Senators. We observe meaningful differences between topics in content, as well as different takes on the same content by partisanship.
    
\end{figure}

Figure \ref{fig:app_topwords} displays the top seven words from each topic with shading indicating how many of the tweets in that topic are sent by Democrats. Topic 1 is Democratic Senators discussing Trump's health care plan, and Topic 3 focuses on climate change. The emerging global pandemic is a focus of several topics, however, each pandemic-related topic captures different aspects of the discussion. Topic 4 focuses on Democrats' efforts to pass a relief bill and Topic 5 is their efforts to improve testing. Topic 17 captures bipartisan efforts to encourage Americans to stay at home and stop the spread. The most Republican coronavirus topic, Topic 29, captures  references to the virus as ``chinese'' and ``communist.'' Our topic model is able to group tweets into coherent units, separating both by subject matter and partisan valance. 

\begin{figure}[ht]
    \centering
    \includegraphics[width=1.0\textwidth]{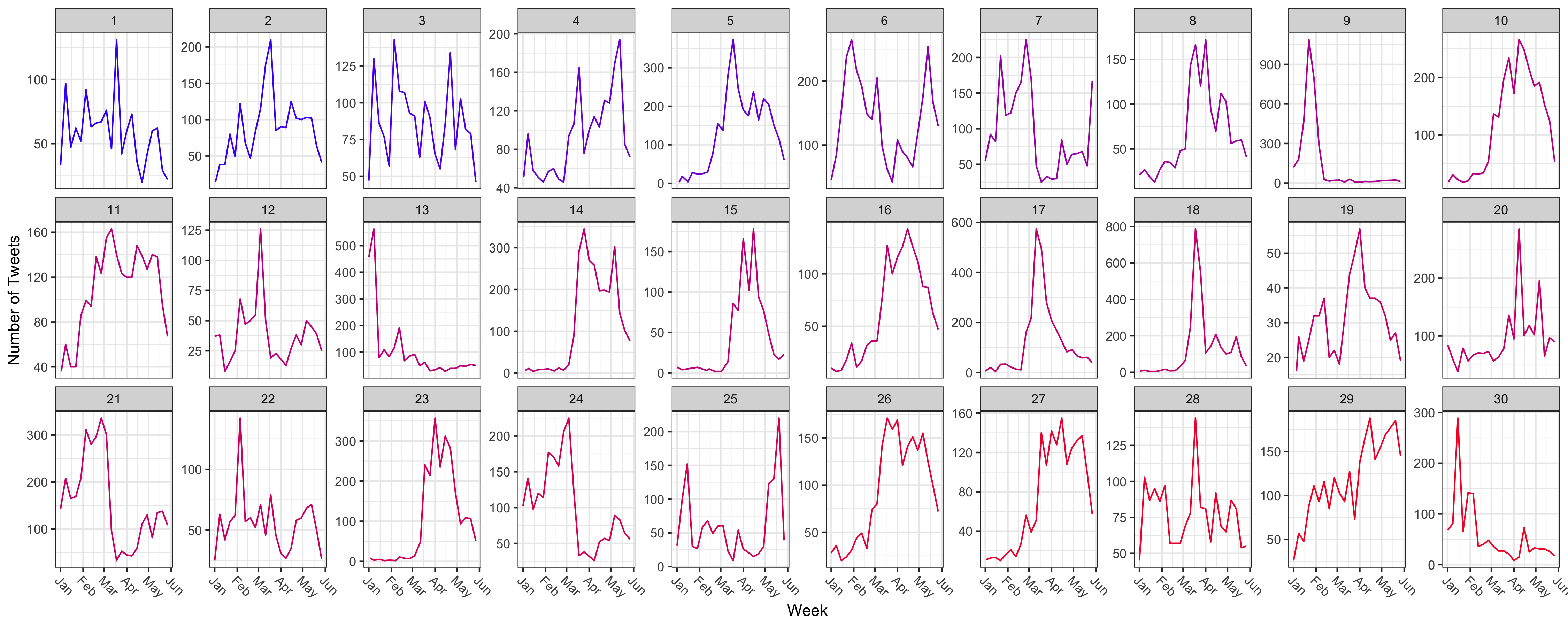}
    \caption{Weekly Topic Counts}
    \label{fig:app_topic_timetrends}
    
    \flushleft
    \footnotesize
    Notes: Each panel shows the weekly number of tweets in each topic over the study period. Topics are ordered and lines are shaded by the percentage of tweets in each topic that are sent by Democratic Senators. Spikes and other time trends indicate that the topic is identifying real variation in how Senators are talking on Twitter. 
\end{figure}

We also examine the changes in the occurrence of topics across time in Figure \ref{fig:app_topic_timetrends}. This gives additional evidence that the top words from each topic are identifying the typical tweets in said topic. Topic 9, the one with the most tweets, has top words indicating that it is about Democrats discussing President Trump's first impeachment trial, which ran from mid-January to early February. Indeed, we see a large spike in that topic in the same time-frame, and almost no discussion afterwards. The coronavirus-related topics, particularly Topic 17, spike mid-March and continue at an elevated level afterwards. Topic 4, Democrats lobbying for passage of legislative items to address the coronavirus, has more tweets from before the pandemic hit because many of the key words overlap with support for passage of other legislative items. Topic 5, focusing on testing, has less general words, so exhibits a more notable increasing trend. The most-partisan topics, 1 and 30, have quite different temporal structures. Topic 1 is tweets from Democrats talking about Trumps lack of a health care policy and defending the Affordable Care Act (ACA), which transition into tweets highlighting the importance of ACA reforms in light of the coronavirus pandemic.

\subsection{Clustering}

We ask our model to find four clusters among the 100 Senators active in this period of the 116th Congress. 

\begin{table}[ht]
    \centering
    \caption{Senators by Cluster and Party (stLDA-C)}
    \label{tab:app_clustertable}
    \begin{tabular}{ccccc}
         & \multicolumn{2}{c}{Party} & \multicolumn{2}{c}{DW-NOMINATE} \\
         \cmidrule{2-3} \cmidrule{4-5}
    Cluster & Democrat & Republican & 1st Dim. & 2nd Dim. \\
\midrule
1 & 14 & 2 & -0.19 & -0.05\\
2 & 31 & 1 & -0.35 & -0.19\\
3 & 0 & 38 & 0.48 & 0.09\\
4 & 2 & 12 & 0.45 & -0.06\\
\bottomrule

    \end{tabular}
    \footnotesize
    \flushleft
Notes: Summary values for each cluster of Senators. Columns 2 and 3 report the number of Senators in each cluster by party. Columns 3 and 4 show the average DW-NOMINATE scores for Senators in each cluster. The differences between Democratic Senators on Twitter appears follow ideological divisions in the party, while the differences in Republican Twitter behavior are not. 
\end{table}

\begin{figure}[htb]
    \centering
    \includegraphics[width=.6\textwidth]{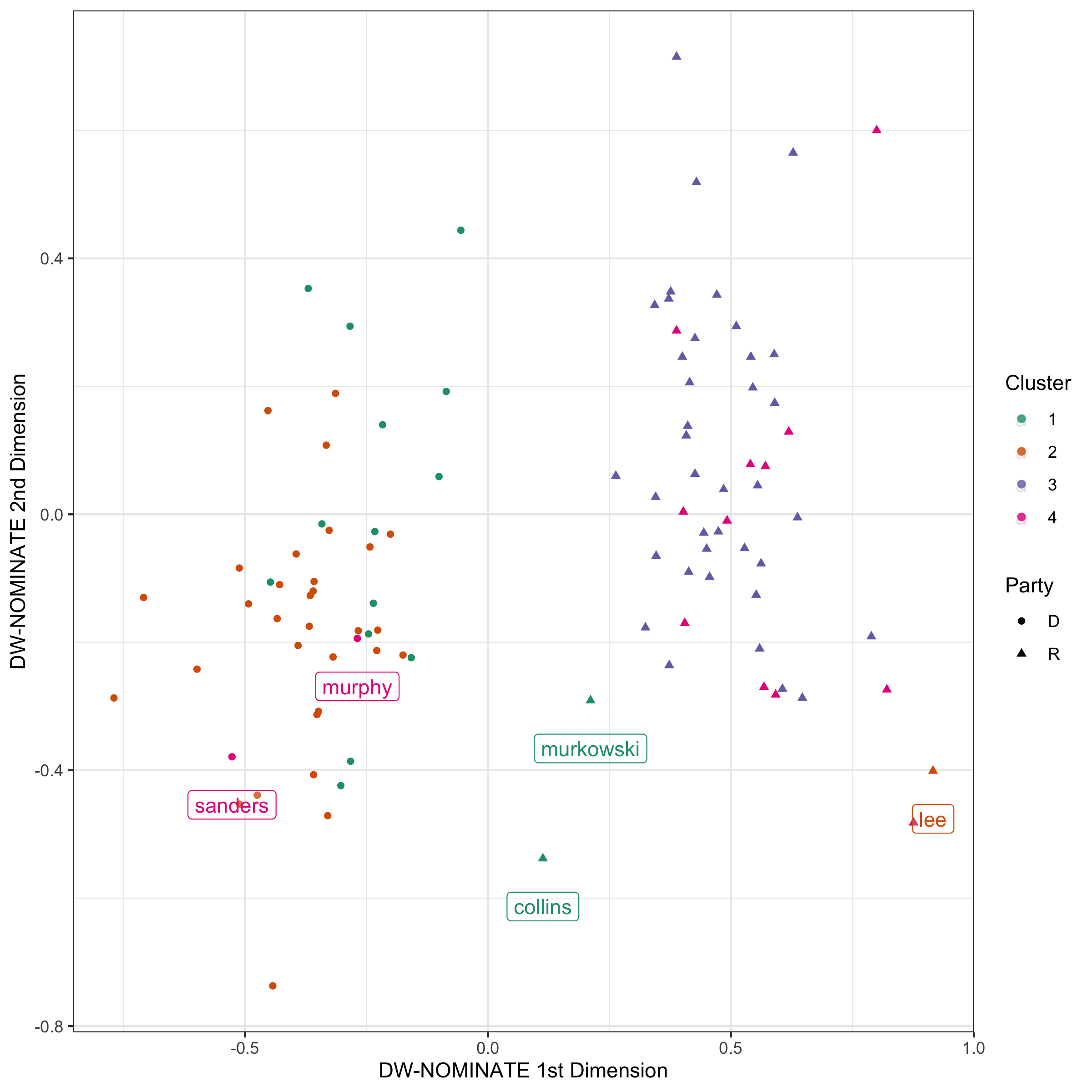}
    \caption{stLDA-C Clusters and DW-NOMINATE Scores}
    \label{fig:app_nominate_clustering}
    
    \flushleft
    \footnotesize
    Notes: Each point represents a Senator in the 116th Congress and their DW-NOMINATE scores, which measure political ideology from voting behavior. Points are colored by which cluster they are classified into by stLDA-C using their Tweets in early 2020. Senators who are classified into a cluster where their party is the minority party are labeled. 
\end{figure}

To assess whether the clustering maps onto ideological differences, we plot the clustering and DW-NOMINATE scores for each Senator in Figure \ref{fig:app_nominate_clustering} and compute party counts and average DW-NOMINATE scores in Table \ref{tab:app_clustertable}. DW-NOMINATE scores are a widely-used latent measure of a politician's ideology learned from their roll call votes. The first dimension, plotted on the x-axis, captures a liberal-conservative split and the second dimension captures other differences, commonly related to social issues \citep{dw_nominate_data}.\footnote{These scores are commonly used in the political science literature, and have even been used as a reference point to evaluate unsupervised modeling of US Senator's Twitter networks \citep{Barbera2015birdsfeather_tweet_together}.} Senators whose party is the minority in their cluster are highlighted.

Cluster 1 contains more Democrats who score higher (more conservative) on the 1st dimension and slightly higher on the 2nd dimension, and cluster 2 contains most of the remaining Democrats. Clusters 3 and 4 split the Republican Party, but do not do so as clearly along the DW-NOMINATE ideology measures. Republican Senators Murkowski and Collins of Alaska and Maine are grouped with the more moderate Democratic Senators in Cluster 1. Republican Senator Lee of Utah is in cluster 2, likely because nearly 20\% of his tweets are estimated to come from topic 6, which contains many tweets from Democrats about President Trump's first impeachment trial, focusing on threats to democracy from foreign interference. Many of the tweets after the impeachment vote are about the re-authorization Foreign Intelligence Surveillance Act (FISA), which was related to impeachment via the use of a FISA court to authorize surveillance of the Trump Campaign. The Democratic Senators clustered with Republicans are Senator Sanders of Vermont and Senator Chris Murphy of Connecticut. Senator Sanders's most common topics are topic 2 and 22, covering 50\% of his tweets. Topic 22 is 80\% Republican tweets that focus on economic issues, which Senator Sanders also highlights.

\begin{figure}[htb]
    \centering
    \includegraphics[width=.6\textwidth]{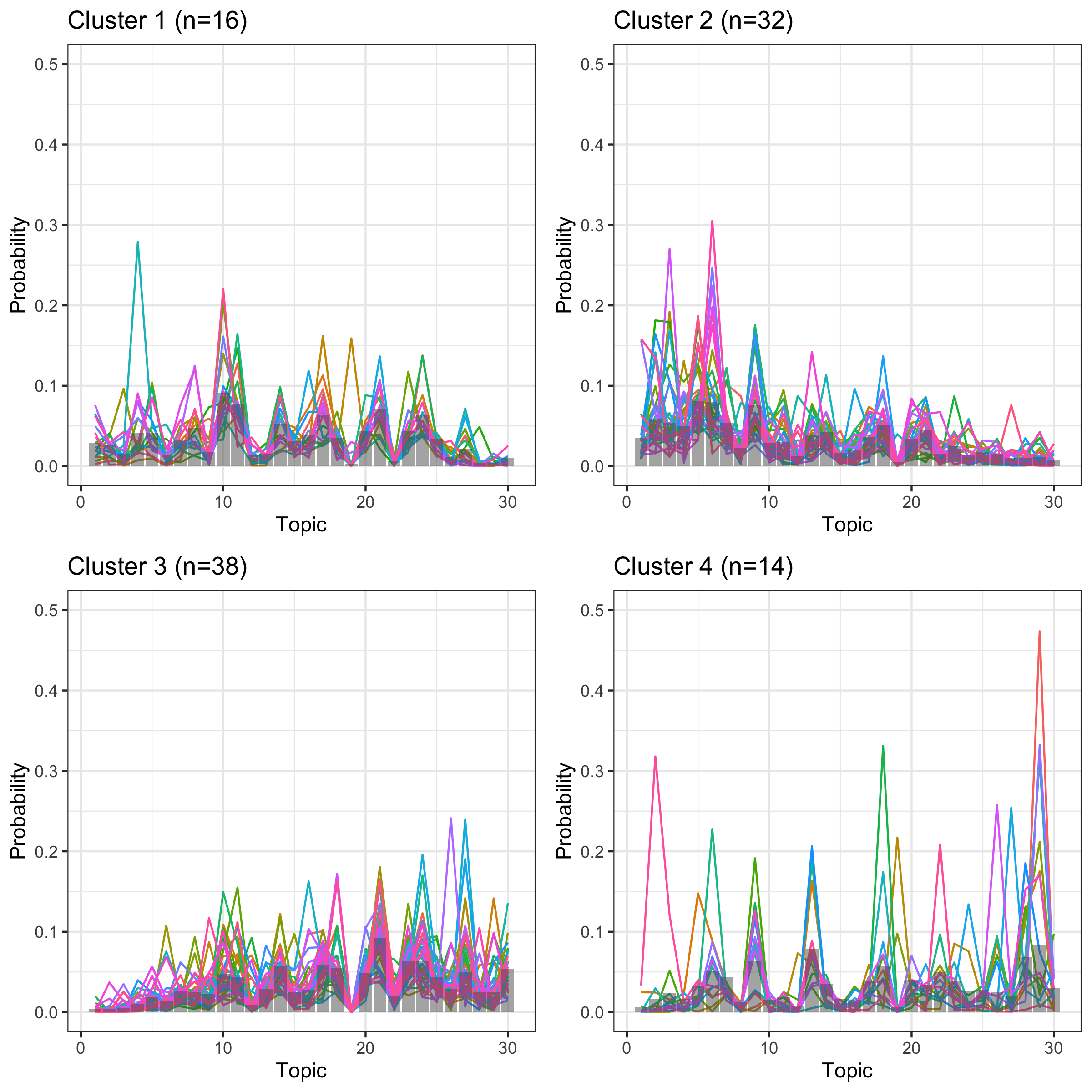}
    \caption{stLDA-C Clusters-Level and User-Level Topic Distributions}
    \label{fig:topic_distributions}
    
    \flushleft
    \footnotesize
    Notes: Each panel shows for a given cluster, all individual-level posterior expected topic distributions as colored lines and the cluster-level expected topic distribution as grey bars. Clusters 1 and 2 are majority Democrats, while 3 and 4 are majority Republicans. Topics are ordered such that Topic 1 has the largest proportion of Democratic tweets and Topic 30 the smallest. 
\end{figure}

To visualize the results, Figure~\ref{fig:topic_distributions} shows all individual users' posterior expected topic distribution and the cluster-level expected topic distributions across all four clusters. Recall that Clusters 1 and 2 are majority Democrats, while 3 and 4 are majority Republicans. Topics are ordered such that Topic 1 has the largest proportion of Democratic tweets and Topic 30 the smallest. The smaller Democratic cluster, Cluster 1, has more frequent users of the mixed topics numbered 10 to 25, which is also reflected in their larger (more conservative) 1st dimension of the DW-NOMINATE score. Cluster 2 Democrats correspondingly concentrate on the lower numbered topics, Topics 1-9 especially, which are more ideologically pure. The Republican clusters are less separated on partisan topics, but rather more on variance. Cluster 3 has much more consistency in topic usage across individuals, while Cluster 4 contains very highly variable distributions. The Senators in Cluster 4 tend to be Republicans with a larger, national profile, such as Senators Mitch McConnel, Ted Cruz, and Rand Paul. Cluster 4 Senators have more than 10 times as many Twitter followers and twice as many tweets in the time period than Cluster 3 Senators, on average. 

This distinction is not surprising given the differences in party organization. Polictical Scientists have documented that the Republican Party has more ideological unity, while the Democratic Party is a coalition of social groups \citep{grossmann2016asymmetric}. As such, Democrats on Twitter separate into a more left-wing and a more moderate group, identified by consistent but disctinct topics in their tweets. Most Republicans, on the other hand, discuss the same set of topics, but a few with national profiles use Twitter to post about a variety of highly variable topics. 

\subsection{Measuring the echo chamber}

Given the focus on echo chambers, we calculate the number of tweets a user would need to read until they have seen at least one tweet on each of the 30 topics to measure insularity across clusters. Specifically, suppose a Twitter user follows only a single account that represents an ``average'' member of a given cluster $g$. The tweets they see are multinomial distributed with probabilities $\alpha_{gt}/\sum_t \alpha_{gt}$. We measure diversity on this Twitter user's timeline by how many tweets do they need to view until they have seen one tweet on every topic. Work on echo chambers usually compares media consumption across predetermined groups, typically partisan affiliation, but our method allows us to measure heterogeneity of echo chambers within party for data-driven groupings of users. 

If $\alpha_{gt}=0$, clearly this number is infinite. Given our estimation method does not compute an exact zero for any cluster-topic pair, we can simulate this finite random variable for each cluster.\footnote{This setup is a variant of the coupon collector problem with unequal probabilities \citep{ferrante2014coupon}. Computing the expectation exactly for 30 topics requires a summation over $2^{30}$ terms and is computationally challenging. As such, we use 1000 Monte Carlo samples that simulate the process to compute estimates and uncertainty measures.} 
This quantity is summarized in columns 2-4 of Table~\ref{tab:topic_coverage}. Clusters 2 and 4 would cover all 30 topics in a little bit less than 300 tweets, Cluster 1 would take almost 500, and Cluster 3 would take almost 1,500. The more liberal Democrats and national-focused Republicans cover the widest variety of topics. The local-focused Republicans in Cluster 3 have the narrowest topic choices, so provide the least diverse coverage in their tweets, reflecting the ideological unity more typical in the Republican Party \citep{grossmann2016asymmetric}. 

This cluster-level ``average'' behavior ignores different user-level variability by cluster, which is quite relevant for the two Republican clusters. As such, we also model a Twitter user who follows $n_g$ accounts all from the same cluster where $n_g$ is the number of Senators in that cluster reported in Table~\ref{tab:app_clustertable}. The topics of tweets they see are simulated by first drawing $n_g$ user-topic distributions from the cluster-specific Dirichlet distribution, then sampling each topic by randomly choosing one of the $n_g$ user-topic distributions to draw a topic from. The topics thus follow a uniform mixture of multinomial distributions. We compute the same statistics and report them in in columns 5-7 of Table~\ref{tab:topic_coverage}. Accounting for this cluster-level variability notably increases the right tail of this distribution, as evidenced in the upper bounds of the 95\% CIs. Accounting for this variability increases the median by about 30 tweets for the more liberal Democrat cluster and almost 150 for the national Republican cluster. Cluster 4 has the highest variability. The sum of the Dirichlet parameters is about 22, while for all other clusters it is around 55. Even though the average Senator in the cluster covers all topics at a similar rate to the liberal Democrats in Cluster 2, the within-cluster heterogeneity often produces user-level distributions that concentrate on small numbers of topics. 

Rows 5 to 10 in Table~\ref{tab:topic_coverage} show the same statistics for combinations of two clusters. The best combination, the one that covers all topics the fastest, is combing either Democratic cluster with Cluster 4 Republicans. These combinations improve topic coverage over following any cluster individually. The median time is reduced by approximately 20\% when following an average Senator and 30\% when following individual Senators. The only inter-party combination that is worse than both intra-party combinations (combinations 1 \& 2 and 3 \& 4) is the moderate Democrats (Cluster 1) and local-focused Republicans (Cluster 3). Even so, the combination is a notable improvement over a user who follows Cluster 3 Republicans alone. 

\begin{table}[ht]
    \centering
    \caption{Expected Number of Tweets until Full Topic Coverage. }
    \label{tab:topic_coverage}
    \begin{tabular}{c ccc c ccc}
                & \multicolumn{3}{c}{Average Senator} & & \multicolumn{3}{c}{Individual Senators} \\
                \cmidrule{2-4} \cmidrule{6-8} 
        Cluster & Median & 95\% CI & Pr(First) & & Median & 95\% CI & Pr(First) \\
        \midrule\midrule
        1 & 488 & $(187,1650)$ & 9\% & & 646 & $(192,5014)$ & 13\% \\
        2 & 281 & $(122,794)$ & 41\% & & 308 & $(130,997)$ & 54\%\\
        3 & 1467 & $(246,7530)$ & 2\% & & 2078 & $(267,83901)$ & 3\% \\
        4 & 266 & $(121,721)$ & 47\% & & 408 & $(145,2877)$ & 30\% \\
        \midrule
        1 \& 2 & 314 & $(128,993)$ & 10\% && 310 & $(114,1204)$ & 16\% \\
        1 \& 3 & 493 & $(137,2454)$ & 7\% && 784 & $(135,6493)$ & 6\% \\
        1 \& 4 & 218 & $(95,824)$ & 29\% && 245 & $(96,1208)$ & 29\% \\
        2 \& 3 & 298 & $(97,1429)$ & 17\% && 324 & $(101,1807)$ & 16\% \\
        2 \& 4 & 213 & $(98,674)$ & 28\% && 236 & $(99,869)$ & 26\% \\
        3 \& 4 & 332 & $(117,1022)$ & 9\% && 536 & $(143,3976)$ & 6\% \\
        \bottomrule
    \end{tabular}
    
    \footnotesize
    \flushleft
Notes: Columns report the number of tweets needed to cover all 30 topics by cluster(s). Columns 2-4 report statistics for a user who follows a single politician who tweets at the cluster-level expected value, i.e. the topics tweets they see are multinomial distributed with probabilities $p_t= \alpha_{gt}/\sum_t\alpha_{gt}$, for rows 1-4, or the mean of two cluster-level expected values for rows 5-10. Columns 5-7 report statistics from accounting for cluster-level variability, a twitter user who follows all politicians from a single cluster, i.e. the topics of tweets they see are a $n_g$-component uniform mixture of multinomial distributions with multinomial probabilities drawn from the cluster-specific Dirichlet parameters $\vect \alpha$ where $n_g$ is the number of politicians in the cluster $g$ listed in Table~\ref{tab:app_clustertable}. The median and 95\% credible intervals are reported. Pr(First) computes the probability that a user who follows politicians from only the given cluster(s) sees all topics first among a cohort of users who each follow only politicians from a unique cluster (rows 1-4) or only politicians from two unique clusters (rows 5-10). 
\end{table}

\subsection{Comparison to Other Methods}

\begin{table}[ht]
    \caption{Summary of Alternative Methods}
    
    \begin{subtable}[h]{\textwidth}
    \centering
    \caption{Senators by Cluster and Party (BTM)}
    \label{tab:app_clustertable_btm}
    \begin{tabular}{ccccc}
         & \multicolumn{2}{c}{Party} & \multicolumn{2}{c}{DW-NOMINATE} \\
         \cmidrule{2-3} \cmidrule{4-5}
    Cluster & Democrat & Republican & 1st Dim. & 2nd Dim. \\
\midrule
1 & 7 & 32 & 0.33 & 0.10\\
2 & 19 & 13 & 0.01 & -0.15\\
3 & 21 & 7 & -0.14 & -0.12\\
4 & 0 & 1 & 0.79 & -0.19\\
\bottomrule

    \end{tabular}
    
    \end{subtable}
    
    \vspace{15pt}
    
    \begin{subtable}[h]{\textwidth}
    \centering
    \caption{Senators by Cluster and Party (Sea-NMF)}
    \label{tab:app_clustertable_seanmf}
    \begin{tabular}{ccccc}
         & \multicolumn{2}{c}{Party} & \multicolumn{2}{c}{DW-NOMINATE} \\
         \cmidrule{2-3} \cmidrule{4-5}
    Cluster & Democrat & Republican & 1st Dim. & 2nd Dim. \\
\midrule
1 & 6 & 12 & 0.21 & 0.10\\
2 & 7 & 23 & 0.29 & 0.06\\
3 & 34 & 17 & -0.06 & -0.15\\
4 & 0 & 1 & 0.79 & -0.19\\
\bottomrule

    \end{tabular}
    \end{subtable}
    
    \footnotesize
    \flushleft
Notes: Summary values for each cluster of Senators with topic modeling using BTM or Sea-NMF and PAM clustering method. Columns 2 and 3 report the number of Senators in each cluster by party. Columns 3 and 4 show the average DW-NOMINATE scores for Senators in each cluster. Clusters are more variable in size and do not map as well onto ideological divisions as clusters identified by stLDA-C. 
\end{table}

\begin{figure}[ht]
    \centering
    \includegraphics[width=.6\textwidth]{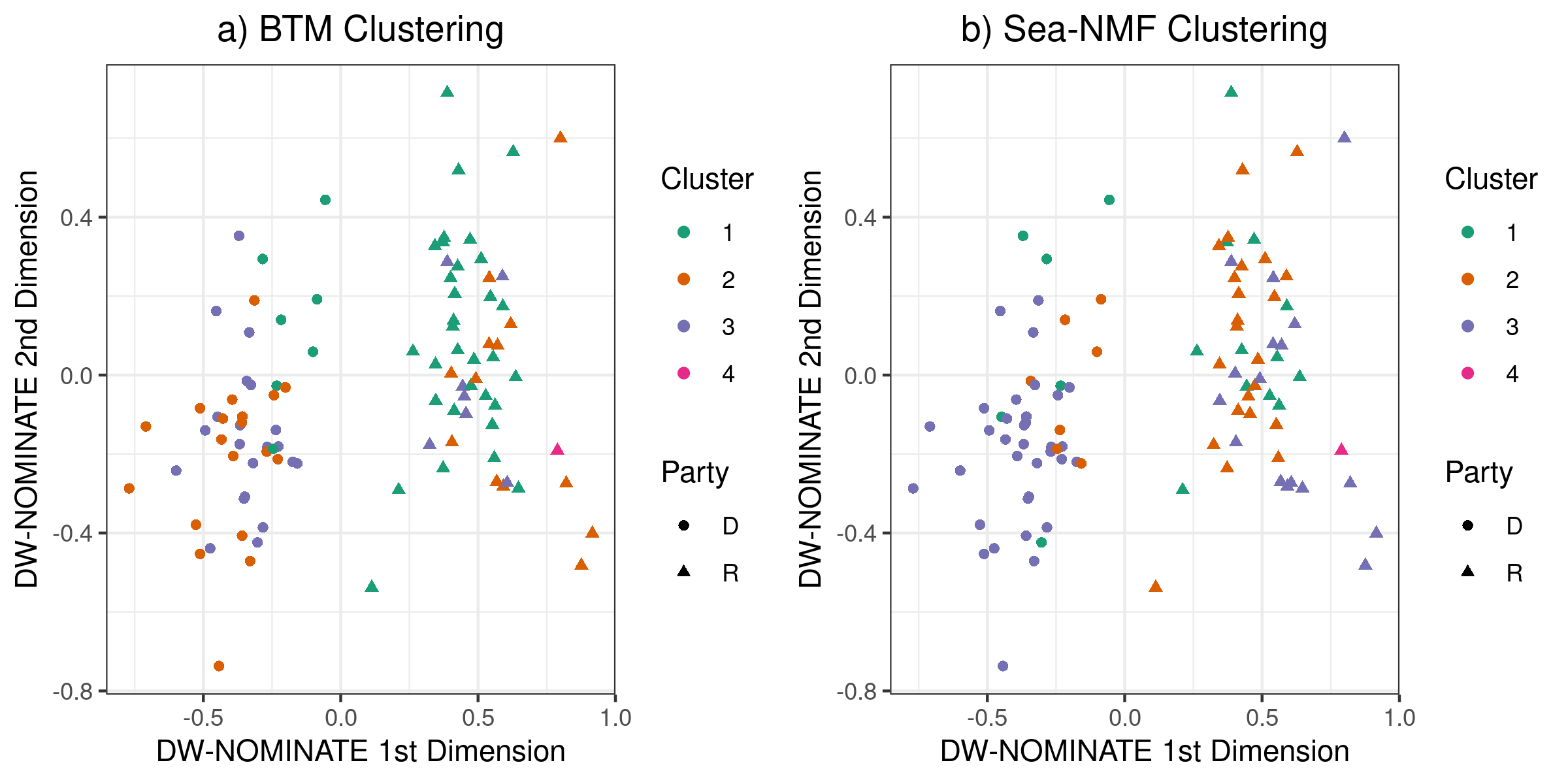}
    \caption{Alternative Method Clusters and DW-NOMINATE Scores}
    \label{fig:app_nominate_clustering_btm_and_seanmf}
    
    \flushleft
    \footnotesize
    Notes: Each point represents a Senator in the 116th Congress and their DW-NOMINATE scores, which measure political ideology from voting behavior. Points are colored by which cluster they are classified into by BTM (panel a) or Sea-NMF (panel b) using their Tweets in early 2020. 
\end{figure}

Next, we construct Table \ref{tab:app_clustertable} and Figure \ref{fig:app_nominate_clustering} using BTM and Sea-NMF. We note that clustering in both of these procedures happens as a second step to topic estimation. Two-step procedures typically do not separate political parties. 

Table \ref{tab:app_clustertable_btm} and Figure \ref{fig:app_nominate_clustering_btm_and_seanmf}a describe the analysis using BTM. We estimate the model with the same number of topics and identify the same number of clusters as in the application of stLDA-C.  Cluster 1 is 32 Republicans and 7 moderate Democrats who score high on the social policy dimension as well.  After that, the clusters are much less clearly based on ideology. Cluster 2 is slightly more Democratic. 

Table \ref{tab:app_clustertable_seanmf} and Figure \ref{fig:app_nominate_clustering_btm_and_seanmf}b describe the analysis using the Sea-NMF method. Again, the clustering does not follow ideological lines as well as the stLDA-C clusters. The majority party in each cluster composes 67\% to 77\% of the members, and 30 Senators are in clusters where their party is in the minority. Using stLDA-C, the smallest majority is 86\% and only 5 Senators are in the minority in their cluster. 

Both two-step clustering procedures place Nebraska Senator Ben Sasse into his own cluster. He sent only 4 tweets from his official Senate account during the time frame, so his topic distribution is very noisily estimated when Tweets are assumed to be independent, as in both the BTM and Sea-NMF methods. Without any shrinkage, given the topics of a user's $n$ tweets, their topic distribution can only be integer multiples of $1/n$ across topics. When $n$ is large, this is not a significant restriction. But when $n$ is small, the potential topic distributions look quite different from users with large $n$ and struggle to capture a continuous parameter like $\theta_u$. In our method, given the topics $\vect Z_u$ and cluster $g_u$, $\theta_u|\vect Z_u,g_u \sim \text{Dir}(\vect Z_u + \vect \alpha_{g_u})$. As the number of tweets increases, the posterior mean shifts from the normalized $\vect \alpha_{g_u}$ values to $\vect Z_u$.  Thus, stLDA-C clusters Senator Sasse with other Republicans rather than clustering him alone because it recognizes that his $\theta_u$ parameter is very noisily estimated and that the few tweets he does send are similar to other Republican tweets. Social media data frequently has this kind of heterogeneity among users, some people post very frequently and others very rarely. Our model handles both types of users well by borrowing strength across multiple levels of data and shrinking noisy estimates towards typical values.

\section{MODEL VALIDATION\label{sec:modelvalidation}}

Tested under known conditions, the model recovers the ground truth for both topic distributions and user clusters better than other methods (where clustering might be performed as a post-estimation step) and produces more coherent topics. Section~\ref{sec:sim_stldac_dgp} uses data simulated from our model, and Section~\ref{sec:alt_dgp} uses data simulated from alternative models where stLDA-C is misspecified. 

We fit stLDA-C using the collapsed Gibbs sampler derived in Section~\ref{sec:collapsed_gibbs}. The closest comparison model is stLDA, single topic LDA without clustering (fit using a collapsed Gibbs sampler). We also compare our model to two more recent methods, BTM, which directly models word-co-occurance \citep{Yan2013biterm}, and the Sea-NMF model, which uses word embeddings learned on the corpus to augment a topic model \citep{Shi:2018:STM:3178876.3186009}.\footnote{The BTM model is implemented in the R package BTM using a Gibbs sampler \citep{btm_r_package}, and the Sea-NMF model is implemented using the replication code provided by the authors on Github.} We also consider tLDA on the full corpus and cLDA on documents created by combining each user's tweets. Standard variational methods were used to estimate these models. We also compare our method to the word mover's distance (WMD) by computing a distance matrix between documents, then running standard clustering algorithms on the results to recover topics. 

We compare each model's ability to accurately recover the true document topics and user clusters, and the coherence of the learned topics. Topic coherence is a standard comparison tool for topic models that measures how frequently high-probability words from the same topic co-occur \citep{mimno2011coherence}. For a given number of highest probability words from each topic, $V$ (15 in our work), the coherence score of the topic is: $\sum_{i=2}^V \sum_{j=1}^{i-1} \log \frac{p(w_i,w_j)+1/D}{p(w_j)}$, where $p(w_i,w_j)$ is the probability of words $w_i$ and $w_j$ co-occurring in the same document and $p(w_k)$ is the marginal probability of word $w_k$ occurring. Smaller values of this score indicate greater coherence. The usual estimates of the probabilities are the co-occurrence or occurrence counts divided by the number of documents $D$. In short texts, this metric is known to be noisy because of the sparsity of word co-occurrences even for high-probability words \citep{Shi:2018:STM:3178876.3186009,quan2015short}. The traditional solution is to estimate the probabilities on an external corpus of longer-documents. To ensure topical similarities, we simulate 100 long documents (100 words) from each user's observed true topic distribution using traditional LDA and calculate the coherence score on that corpus.

\subsection{Simulation Study\label{sec:sim_stldac_dgp}}

We simulate data according to the generative model for single topic LDA with clustering (stLDA-C). To simulate realistic topic distributions, we estimated 10 word-topic distributions over 4,534 unique words from the Guardian news corpus included in the quanteda R package and used the posterior mean as the ``true'' topic distributions for our simulations \citep{quanteda}. We then simulated data from 4 clusters with 10 users per cluster. We choose parameters for the simulation such that there are both similar and distinct clusters: $\alpha_{1t} = 10$ for all topics, $\alpha_{2t} = 20$ for $t \leq 5$ and 0 otherwise, $\alpha_{3t} = 20$ for $t>5$ and 0 otherwise, and $\alpha_{4t} = 25$ for $t\in\{2,3,4,5\}$ and 0 otherwise. $\theta_u$ in clusters 2 and 3 will place positive probability on only disjoint sets of five topics. Cluster 4 is extremely similar to cluster 2; it only differs by one topic. We draw 100 tweets of 13 words each per user, for a total of 4,000 documents.\footnote{The average tweet in the data application has 13 words.} Note that while the data generating process is the stLDA-C model, the parameters lie outside of the prior with 0 values in certain cluster-level parameters. We use diffuse, non-informative priors of $\eta=\nu=1$ and $\alpha_g = \vect{m_g} c_g$ with $\vect{m_g} \sim Dir(1)$ and $c_g \sim N(100,50)$.\footnote{The same priors were used in Section~\ref{sec:vb_gibbs_comparison}.}

\textbf{Coherence.} Our model produces the most coherent topic distributions. Table \ref{tab:topic_coherence} shows the topic coherence scores averaged across topics for each applicable model. The true distributions are obviously the most coherent, but our method and stLDA without clustering are not far behind. Sea-NMF and BTM significantly outperform cLDA and tLDA. The alternative simulation models are discussed in the next section. 

\begin{table}[bt]
\caption{Topic Coherence}\label{tab:topic_coherence}
\vskip 0.15in
\begin{center}
\begin{small}
\begin{tabular}{lccc}
\hline
Estimation & \multicolumn{3}{c}{Simulation Model} \\
\cmidrule{2-4}
    \multicolumn{1}{c}{Method} &  stLDA-C & stLDA & cLDA \\
    \hline
    cLDA & -364.02 & -365.57 & -362.74 \\
    tLDA &  -358.49 & -368.49 & -363.96 \\
    BTM & -270.47 & -260.89 & -269.18\\
    Sea-NMF & -268.13 & -267.80 & -318.65\\
    stLDA & -256.20 & -268.61 & -309.87\\
    stLDA-C & -254.59 & -259.47 & -242.74\\
    \hline
    True & -249.56 &-253.61 &-253.83 \\
    \hline
\end{tabular}
\end{small}
\end{center}

     \vspace{.25cm}
    \footnotesize
    \flushleft
    Notes: Topic coherence \citep{mimno2011coherence} is calculated as $\sum_{i=2}^V \sum_{j=1}^{i-1} \log \frac{p(w_i,w_j)+1/D}{p(w_j)}$, where $p(w_i,w_j)$ is the number of documents where words $w_i$ and $w_j$ both occur divided by the number of documents $D$. This metric is calculated on an external corpus of long documents generated from the same user-topic distributions. ``True" uses the true topic distributions. We find that our method produces the most coherent topics across all estimation methods and simulation conditions. 
\end{table}

\textbf{Topic recovery.} The first column of Table \ref{tab:dt_recovery} compares document topic recovery for our method and comparison methods under the stLDA-C generative model. Each iteration of the Gibbs sampler provides a sample of $Z_{ud}$, so the proportion of times $Z_{ud} = t$ is a posterior estimate of $P(Z_{ud} = t|\vect{W})$. Documents are classified into the highest posterior-probability topic. Clustering is evaluated with normalized mutual information (NMI), which measures the mutual information between the estimated and ground truth clusters normalized to be between 0 (no information shared) and 1 (perfect correlation). 
The more novel BTM and Sea-NMF methods are able to recover many topics accurately, but they are notably worse than the single-topic models. stLDA without clustering is able to recover document topics with comparable accuracy to stLDA-C.

\begin{table}
    \caption{Document Topic Recovery (NMI)}\label{tab:dt_recovery}
    \begin{center}
    \begin{small}
    \begin{tabular}{lcc}
    \hline
  Method & stLDA-C DGP & stLDA DGP \\
  \hline
    tLDA &  0.052 & 0.023 \\ 
    WMD & 0.247 & 0.233 \\ 
    BTM & 0.890 & 0.878 \\ 
    Sea-NMF & 0.925 & 0.853 \\ 
    stLDA &  0.952 & 0.904 \\ 
    stLDA-C & \textbf{0.955} & \textbf{0.953} \\ 
   \hline
    \end{tabular}   
    \end{small}
    \end{center}
    %\vspace{.25cm}
    
    \flushleft
    \footnotesize
    Notes: Topic recovery is evaluated with normalized mutual information, a measure of cross-entropy from the estimated and true labels. An NMI of 1 indicates perfect recovery of the ground truth. For topic models, documents are classified by the highest posterior probability topic. For WMD, the partitioning around medoids (PAM) clustering method is used, a more robust version of k-means \citep{Reynolds2006}. stLDA-C recovers nearly every document's topic, while WMD and tLDA are not much better than random guesswork. 
\end{table}

\textbf{Cluster recovery.} Our approach has the best cluster recovery with each user being mapped to the correct cluster. Table \ref{tab:cluster_recovery} shows the results. To highlight the places other methods fail in the clustering, we show cluster-specific true- and false-positive rates rather than a global metric. The other methods must rely on a two-stage cluster estimation procedure: user-topic distributions are calculated as the proportion of each users' tweets that are classified into each topic, then those distributions are passed to the PAM clustering algorithm, a more stable variant of k-means clustering \citep{Reynolds2006}. The principal confusion is between clusters 2 and 4, which only differ by one topic. All alternative models identify that clusters 2 and 4 differ on topic 1, but some users from cluster 2 who talk about topic 1 less than the average user are grouped into cluster 4. The independent clustering necessarily does not leverage the correct distance metric between topic distributions. Only our method stLDA-C, which unifies the estimation of clusters and topics, is able to recover the clusters with 100\% accuracy because it recognizes the differences across topic 1 are the most significant differences between clusters 2 and 4. 

\begin{table}
    \caption{Cluster Recovery from stLDA-C Simulated Data}\label{tab:cluster_recovery} 
    \begin{center}
    \begin{small}
    \begin{tabular}{ccccccccccccccc}
      \hline
      True & \multicolumn{2}{c}{WMD} & \multicolumn{2}{c}{BTM} & \multicolumn{2}{c}{Sea-NMF} & \multicolumn{2}{c}{stLDA}  & \multicolumn{2}{c}{stLDA-C}  \\
     Cluster  & TPR & FPR & TPR & FPR & TPR & FPR & TPR & FPR & TPR & FPR \\
      \hline
          1 & 100\% & 3\% & 100\% & 0\% & 100\% & 0\% & 100\% & 0\% & 100\% & 0\% \\ 
        2 & 60\% & 23\% & 70\% & 0\% & 70\% & 0\% & 70\% & 0\% & 100\% & 0\%  \\ 
        3 & 100\% & 0\% & 100\% & 0\% & 100\% & 0\% & 100\% & 0\% & 100\% & 0\% \\ 
        4 & 70\% & 20\% & 100\% & 10\% & 100\% & 10\% & 100\% & 10\% & 100\% & 0\% \\ 
       \hline
    \end{tabular}
    \end{small}
    \end{center}
    
    \footnotesize
    \flushleft
    Notes: TPR for cluster $c$ is the largest proportion of users with cluster $c$ mapped to a single estimated cluster. FPR is the proportion of users with that estimated cluster among users in cluster $c'\neq c$. For stLDA-C, users are classified into the highest posterior probability cluster. For other methods, user-level topic distributions are passed to the the PAM clustering method to identify 4 clusters \citep{Reynolds2006}. stLDA-C recovers all clusters accurately, while two-step procedures struggle to separate clusters 2 and 4. Cluster 4 contains users who talk about all but one of the topics used by users in cluster 2.  
\end{table}

\begin{figure}[ht]
    \begin{center}
    \includegraphics[width=.8\textwidth]{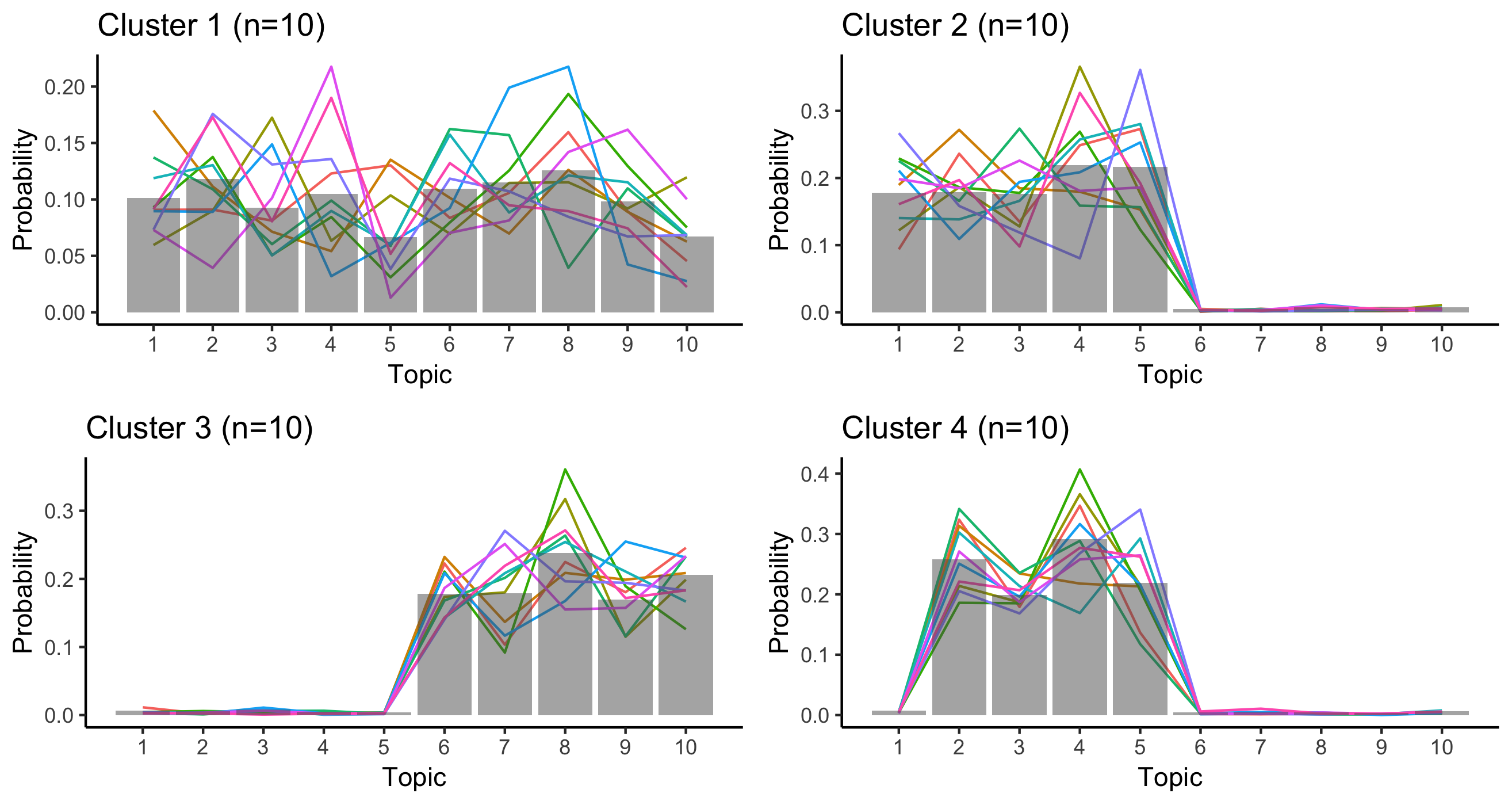}
    \caption{User Topic Distributions by Cluster (Simulated Data)}
    \label{fig:groundtruth_usertopics}
    \end{center}
    
    \footnotesize
    \flushleft
    Notes: Cluster-level expected values are shown in grey. Each user's estimated topic distribution is shown with a colored line. stLDA-C correctly identifies which topics each cluster discusses, capturing both the cluster-level averages and user-level variability. 
\end{figure}

A useful visualization of the results compares the expected topic distribution of each cluster and the variability of users within a cluster. Figure \ref{fig:groundtruth_usertopics} shows each estimated user topic distribution on top of the cluster mean. Our model recovers the ground truth and captures the fact that there are four groups of users with similar but not identical topic distributions.

\subsection{Model Miss-Specification\label{sec:alt_dgp}}

In the above simulations, we generated data according to the stLDA-C generative model. In this section, we simulate data according to two alternative models and test whether our method still performs well. We preserve the size of the simulated data from simulations for stLDA-C to allow comparison of results across simulations. 

The first alternative simulation model is simple stLDA, a submodel of stLDA-C. Effectively, data are simulated from only one cluster. Each $\theta_u$ is sampled from a uniform distribution over the 10 dimensional simplex, then 100 topics are sampled from $\theta_u$, and one document for each sampled topic is generated. We compare the same models in the last section in terms of document-topic recovery, user-cluster recovery, and topic coherence. 

We use the same parameter settings for our comparison models. Notably, we allow stLDA-C to use the same maximum number of clusters (four), and discover that the model correctly identifies that only one cluster is needed. Other comparison models rely entirely on a researcher-specified number of clusters, rather than specifying a maximum number and letting the data identify the true number of clusters needed. 

Table \ref{tab:topic_coherence} shows the topic coherence scores for each model in this simulation in the second column. stLDA-C and BTM are again quite similar, with our model estimating slightly more coherent topics. Column 2 in Table \ref{tab:dt_recovery} shows the NMI for learned document topics for each method under stLDA simulations. All four of the most advanced methods have high NMI values. Our model has the highest of the four, even higher than the correctly-specified stLDA method. This simulation shows that our setup is able to adapt to sub-models.

Next, we simulate data from the cLDA generative process by generating one long document for each user, then breaking it up into short, tweet-length documents. Thus, our model is miss-specified since it assumes a single topic for each document, whereas the truth is that multiple topics are present. We generate user-specific topic distributions with the same clustering as in the stLDA-C simulations: one cluster discusses all topics, one cluster only half of the topics, another cluster discuss the other half, and the final cluster discusses all but one of topics another cluster uses. We compare all models on cluster recovery and topic coherence. 

Our model correctly identifies that clusters 1 and 3 are distinct, but merges the clusters that differ on only one topic into a single cluster. Only three of the maximum of four clusters have users in them. Other methods are able to somewhat replicate this recovery, but none are as good as stLDA-C. They confuse outliers in clusters two and four with members of the other cluster and include some members of cluster 1, who discuss all topics, in the combined clusters of 2 and 4. The results are shown in Table \ref{tab:cr_cLDA}. 

Our model also performs best in terms of topic coherence. We again simulate many long documents from the same user-topic distributions to compare the quality of estimated topics. Column 3 in Table \ref{tab:topic_coherence} shows that our method even marginally out-performs the true topic distributions in terms of coherence. 

\begin{table}
    \caption{Cluster Recovery from cLDA Simulated Data}\label{tab:cr_cLDA} 
    \begin{center}
    \begin{tabular}{ccccccccccc}
  \hline
  True & \multicolumn{2}{c}{WMD} & \multicolumn{2}{c}{BTM} & \multicolumn{2}{c}{Sea-NMF} & \multicolumn{2}{c}{stLDA}  & \multicolumn{2}{c}{stLDA-C}  \\
 Cluster  & TPR & FPR & TPR & FPR & TPR & FPR & TPR & FPR & TPR & FPR \\
  \hline
  1 & 70\% & 7\% & 90\% & 0\% & 100\% & 0\% & 90\% & 0\% & 100\% & 0\% \\ 
    2 & 100\% & 20\% & 100\% & 37\% & 50\% & 10\% & 100\% & 37\% & 100\% & 33\% \\ 
    3 & 100\% & 3\% & 50\% & 0\% & 100\% & 0\% & 60\% & 0\% & 100\% & 0\% \\ 
    4 & 40\% & 40\% & 100\% & 37\% & 70\% & 17\% & 100\% & 37\% & 100\% & 33\% \\ 
   \hline
\end{tabular}
\end{center}

\flushleft\footnotesize
TPR for cluster $c$ is the largest proportion of users with cluster $c$ mapped to a single estimated cluster. FPR is the proportion of users with that estimated cluster among users in cluster $c'\neq c$. For stLDA-C, users are classified into the highest posterior probability cluster. For other methods, user-level topic distributions are passed to the the PAM clustering method to identify 4 clusters \citep{Reynolds2006}. stLDA-C has the best cluster recovery. It separates users into three clusters. successfully separating all members of clustrs 1 and 3, and it combining clusters 2 and 4 which only differ on one topic. All other procedures fail to fully separate 1 and 3 and do not correctly identify the similarities among clusters 2 and 4. 
\end{table}

\section{DISCUSSION}

Social media data and short text more broadly are becoming an important area of study for researchers. Standard models for long text already have to address issues regarding high dimensionality when treating words as features. Models for short text have to address those same issues with the added complication of increased sparsity in observed word co-occurrences and frequent desire to use model results in a downstream task. stLDA-C solves all of those issues by linking the topics of words used in the same short text, borrowing strength across documents generated by the same user, and clustering users with a hierarchical prior that accounts for uncertainty in user-level parameters. 

Our model is able to recover topics and clusters of users in simulated data, whereas other state-of-the-art methods struggle with one or both tasks. We developed the unified estimation procedure for topics and clusters because of the frequent desire to categorize at both the post and user level simultaneously. Our variational approximation allows our model to scale to large corpora typical of social media data. As demonstrated in the application, those user-level groupings are often related to important user features such as political party. A further extension of our algorithm could use the information on these features in the cluster estimation itself, such as in \citet{roberts2013structural}. We use only topic frequencies to group users, but one could build upon our clustering step models that leverage demographic information as well. 

Even without including external information about users, our application demonstrates that such information can be gleaned from the clustering. We capture distinct partisan topics and group US Senators into clusters that reflect partisan ideology. Our model separates tweets about several different aspects of the coronavirus pandemic, and identifies distinct subgroups within each party that tweet differently. The novel measure of the echo-chamber characterizes the degree of this separation and has important ramifications for understanding the scope and character of exposure to cross-partisans on social media. To learn what the other side really thinks about current events, citizens will need to actively seek out discussion from leaders in the opposing party.

\clearpage
\bibliographystyle{ba}
\bibliography{master}

\clearpage

\appendix

\input{appendix_input}

\end{document}

%% file: Floats/hstLDA_phi.tex
% model_lda.tex
%
% Copyright (C) 2010,2011 Laura Dietz
% Copyright (C) 2012 Jaakko Luttinen
%
% The MIT License
%
% See LICENSE file for more details.

% Latent Diriclet allocation model

%\beginpgfgraphicnamed{model-lda}
% \begin{tikzpicture}[x=1.7cm,y=1.8cm]
\begin{tikzpicture}[x=1.5cm,y=1.3cm]

  % Nodes

  \node[obs]                   (W_u)      {$\mathbf{W}_{ud}$} ; %
  \node[latent, above=of W_u,yshift=-.3cm]    (Z_u)      {$Z_{ud}$} ; %
  \node[latent, above=of Z_u,xshift=-.8cm,yshift=-.3cm]    (theta_u)  {$\theta_u$}; %

  %\node[obs, right=of W_u]      (W_v)      {$\mathbf{W}_v$} ; %
  %\node[latent, above=of W_v]    (Z_v)      {$Z_v$} ; %
  %\node[latent, above=of Z_v]    (theta_v)  {$\theta_v$}; %

  \node[latent, above=of theta_u,yshift=-.3cm] (atheta) {$\alpha_g$};
  
  \node[const,left=of atheta]  (alpha) {$\alpha$};

  \node[latent, right=of theta_u,xshift=-1cm] (g) {$G_u$};
  \node[latent, right=of g,xshift=0cm] (phi) {$\phi$};
  \node[const, below=of phi,yshift=-.3cm] (nu) {$\nu$};
  % Factors
  %\factor[above=of X]     {X-f}     {Multi} {} {} ; %
  %\factor[above=of T]     {T-f}     {left:Multi} {} {} ; %
  %\factor[above=of theta] {theta-f} {left:Dir} {} {} ; %

  % More nodes
  \node[latent, left=of W_u] (beta)  {$\beta_t$}; %
  \node[const, above=of beta]  (aphi) {$\eta$}; %

  %\factor[above=of phi] {phi-f} {right:Dir} {} {} ; %
  
  \edge {nu} {phi}
  \edge {phi} {g}
  \edge {g} {theta_u}
  
  \edge {alpha} {atheta}
  
  \edge {atheta} {theta_u} ; %
  \edge {theta_u}  {Z_u} ; %
  \edge {Z_u}       {W_u}
  \edge {beta}    {W_u}; %
  
  %\edge {atheta} {theta_v} ; %
  %\edge {theta_v}  {Z_v} ; %
  %\edge {Z_v}       {W_v}
  %\edge {beta}    {W_v}; %
  
  \edge {aphi}   {beta}   ; %

  %\gate {X-gate} {(X-f)(X-f-caption)} {T}
  
  \plate {} { %
    (beta) %
  } {$\forall t \in \mathcal{T}$} ; 

  \plate {plateDu} %
  {(Z_u) (W_u)} %
  {{$\forall d \in \mathcal{D}_u$}}
  
  %  \plate {plateDv} %
  %{(Z_v) (W_v)} %
  %{{$\forall d \in \mathcal{D}_v$}}
  
  \plate{plateu}{(plateDu)(theta_u)(g)}{$\forall u : G_u=g$}
  
  %\plate{platev}{(plateDv)(theta_v)}{}
  
  \plate{}{(plateu)(atheta)}{$\forall g \in \mathbf{G}$}

\end{tikzpicture}
%\endpgfgraphicnamed

%%% Local Variables: 
%%% mode: tex-pdf
%%% TeX-master: "example"
%%% End: 

%% file: other_methods.tex
\textbf{Introducing a user clustering layer.} On Twitter, researchers frequently want to not only learn topics of tweets, but also group users into categories. Previous work has estimated user-level topic frequencies, and then used those estimates in a separate clustering procedure. Our method links the topic modeling and clustering task, allowing full propagation of uncertainty and combining the two inference steps. This clustering layer is reminiscent of the document clustering in tLDA due to \citet{wallach2008structured}, but because there are many documents per user, the clustering is more naturally described on the user layer. Each user $u$'s cluster is denoted by $G_u$. For users in the same cluster $g$, $\theta_u \sim \text{Dir}(\alpha_g)$. This hierarchical model allows for shrinkage of noisily estimated $\theta_u$ parameters from users who post infrequently to a common mean and borrowing strength for that estimation across users with similar topics. This is especially important when there is large heterogeneity in the number of posts from each user, which we demonstrate in the application. 

Using multiple clusters, rather than a single hierarchical prior, is important in the context of social media. The scale and structure of social media data often means that many users discuss completely distinct sets of topics. Literature on echo chambers on social media platforms document this phenomenon \citep{Bakshy2015}. Having a single hierarchical prior, where all $\theta_u$ are drawn from the same distribution, will shrink estimates of $\theta_u$ to an average taken across the entire population of users. With multiple clusters, noisily estimated $\theta_u$ parameters are shrunk towards the average of users they are most similar to.

Our model treats a copora as a mixture of a finite number of clusters and finite number of topics, both of which are assumptions that have been relaxed in the Bayesian non-parametric literature using Dirichlet processes to allow for countably infinite numbers of mixture components. The multi-corpora Heirarchical Dirchlet Process model  of \cite{teh2006hierarchical}, for example, can be thought of as an infinite dimensional extension of our model when cluster labels (corpora in their terminology) are fixed and known.\footnote{A large literature on discrete, supervised LDA has similar structural form to our model here when class or cluster labels are observed, such as the recent work \cite{wang2021bayesian}. These models, while similar in form, are not directly comparable to our model because we must infer the clustering from the data. Dirichlet process models that use latent clustering, as in the Nested Dirichlet Process \citep{rodriguez2008nested}, do not allow sharing of atoms across clusters. In our case, that would mean all topics are cluster-specific, which is not desirable because we want to compare topic choice across clusters.} As described below, the scenarios and data we designed our model to analyze require a smaller number of interpretable topics, so allowing for large numbers of topics is not necessarily an improvement. Moreover, when the subjects of individual tweets are of particular interest and domain experts can assess meaningfulness of topics, as in our application, human validation of topic quality and utility for downstream tasks is often preferable to the data-driven selection in the Hierarchical Dirichelt Process model. 

\textbf{Connections to single topic LDA.} Our model naturally generalizes the two different single topic per document models, Dirichlet-Multinomial Mixture (DMM) and single topic LDA (stLDA). In the DMM model, each short text is drawn from a single latent topic and the corpus has a single set of mixture proportions \citep{nigam2000text}. In this model, each document belongs to a single latent topic and words in the document are a multinomial draw from a cluster-specific probability distributiton over words. Several estimation methods have been developed, including EM and collapsed Gibbs sampling \citep{nigam2000text,Yin:2014:DMM:2623330.2623715}. DMM is a simpler version of the model proposed here (and by \cite{zhao2011comparing}). Indeed, the DMM model can be recovered by removing the user- and cluster-specific plates in Figure \ref{fig:stLDA_graph}. 

stLDA from \citet{zhao2011comparing} overcomes the sparsity of word co-occurances by ensuring that all words in a document contribute to the estimation of the word distribution of a single topic, $t$. By placing the topic mixing at the user rather than document level, stLDA captures dependence among documents by the same user and improves topic estimation by borrowing information across these documents. Topics of documents by the same user are conditionally independent given the user's topic distribution $\theta_u$, but become dependent after integrating $\theta_u$ out, whereas in tLDA the topics of individual documents remain independent when each document's topic mixture is integrated out. To learn the topic of a given tweet by user $u$, stLDA places significantly more weight on words and documents produced by $u$ because they inform beliefs about $\theta_u$. This is in contrast to tLDA in which all words are given equal importance, regardless of the user who produced them. stLDA can be recovered from our model by integrating over cluster parameters $\alpha_g$, $G_u$, and $\phi$ in Figure~\ref{fig:stLDA_graph}.

\textbf{Connections to combined LDA.} The way many researchers apply topic models to Twitter data is to combine all documents by the same user into one large document and estimate tLDA models on the smaller set of longer documents \citep{hong2010empirical,pennacchiotti2011machine,steinskog-etal-2017-twitter,barbera2018leads}. We call this method cLDA. The topics of words in the same tweet should be related, even given a user's average topic frequency, because tweets are often focused on a narrow subject due to character limits. However, because cLDA uses a simple bag-of-words model, the information that certain words were used in different tweets is lost.  In cLDA, the topics of words in the same tweet by one user are conditionally independent given the user's topic frequency. Suppose words $w$ and $w'$ appear in the same tweet by user $u$ and they have potentially distinct topics $Z_w$ and $Z_{w'}$. Conditional on $\theta_u$, the user's topic distribution, knowing the topic of $w'$ provides no additional information about the topic of $w$. $P(Z_{w}=t|Z_{w'}=t,\theta_u) = P(Z_{w}=t|\theta_u) = \theta_{ut}$, the marginal probability that user $u$ picks topic $t$. This is true for all words authored by a single user, regardless of whether they appeared in the same tweet. This result does not capture how most users create posts. Suppose a politician talks about national politics half of the time and local issues the other half. Knowing that one of the words in a tweet is about national politics should increase the belief that other words in the tweet are also about national politics, even knowing the user spends on average half the time on each issue. cLDA resolves the issue of short documents by assuming a level of conditional independence that fails to capture the document generating process. 

In stLDA, the topics of words in the same tweet are identical, which more accurately captures how tweets are created. This means that, in stLDA, knowing the topic of one word in a tweet identifies the topics of all other words. In the same situation as above, where words $w$ and $w'$ appear in the same tweet by user $u$, $Z_w=Z_{w'}=Z_{ud}$. Thus, conditional on $Z_{w'}$, $Z_w$ is independent of $\theta_u$. $P(Z_{w}=t|Z_{w'}=t,\theta_u) = 1$. {\it One can thus think of cLDA and stLDA as two extremes for modeling the dependence of topics of words in the same tweet, conditional on the user's average topic selection}. stLDA still allows for borrowing strength across the topics of the user's other tweets. $\theta_u$ is learned from those topics, which strongly influences the topic probabilities of new tweets.

DMM models have also been augmented with word embeddings to improve topic estimation by identifying semantically similar words \citep{li2016gpudmm}. Again, these models rely on either a pre-trained set of embeddings or an auxiliary dataset known to overlap, at least partially, in subject matter. We feel that these are quite strong assumptions, particularly given the target application of our model is social media data where semantic content is different from longer text. A common application of short text models is to news headlines, which have a more plausible link to longer text documents, but our model is designed for a different area of application.

%% file: appendix_input.tex
\section{Collapsed Gibbs Sampler}

Proof for the collapsed Gibbs sampler. $Z_{ud}$ is the topic of tweet $d$ by user $u$. $\theta_u$ is the user's topic distribution, $\beta_t$ is the topic distribution $t$ over words. We desire $P(Z_{ud}=t|\vect{Z_{-ud}},\vect{W})$, the distribution of $Z_{ud}$ given the topics of all tweets except tweet $d$ by user $u$ and the word counts of all tweets $\vect{W}$. Given the topic of a tweet $Z_{ud} = t$, the word counts $\vect{W_{ud}}$ are a multinomial draw from topic distribution $\beta_t$. $\theta_u$ identifies the unconditional, prior distribution for $Z_{ud}$. Via Bayes' rule, the full conditional for $P(Z_{ud}=t|\theta_u,\beta_t,\vect{Z_{-ud}},\vect{W})$ proportional to the product of this multinomial likelihood and $\theta_u$ prior. Thus, given $\beta_t$ and $\theta_u$, the posterior for $Z_{ud}$ does not depend on any words or topics besides those in tweet $ud$. 

\[P(Z_{ud}=t|\theta_u,\beta_t,\vect{Z_{-ud}},\vect{W}) \propto P(\vect{W_{ud}}|Z_{ud}=t,\beta_t)P(Z_{ud}=t|\theta_t) \propto \prod_i \beta_{ti}^{W_{udi}} \theta_{ut}\]

From the paper: 

\begin{align*}
    p(&Z_{ud} = t|\vect{Z_{-ud}},\vect{W}) \\
    &= p(Z_{ud} = t|\vect{Z_{-ud}},\vect{W_{ud}},\vect{W_{-ud}}) \\
    &\propto p(\vect{W_{ud}}|Z_{ud}=t,\vect{Z_{-ud}},\vect{W_{-ud}})p(Z_{ud}=t|\vect{Z_{-ud}},\vect{W_{-ud}})\\
    &= \int p(\vect{W_{ud}}| Z_{ud}=t,\beta_t) p(\beta_t | \vect{Z_{-ud}},
    \vect{W_{-ud}}) \ d\beta_t \times \\
    &\hspace{12pt} \int p(Z_{ud}=t | \theta_u) p(\theta_u|\vect{Z_{-ud}}) \ d\theta_u\\ 
    &= \int \prod_i \beta_{ti}^{W_{udi}} p(\beta_t|\vect{Z_{-ud}},\vect{W_{-ud}}) \ d\beta_t \times \\
    &\hspace{10pt} \int \theta_{ut}p(\theta_u|\vect{Z_{-ud}})\ d\theta_u 
\end{align*}

Conditional on $G_u=g$ and $\alpha_g$, the unconditional distribution of $\theta_u$ is: $p(\theta_u) \sim \text{Dir}(\alpha_g)$. The Dirichlet is conjugate with multinomial data. Let $\vect{Z_{-d}^{(u)}}$ denote the topic counts of user $u$ excluding the topic of tweet $d$. 

\begin{align*}
   p(\theta_u|G_u=g,\alpha_g,\vect{Z_{-ud}}) &\propto p(\vect{Z_{-d}^{(u)}}|\theta_u) p(\theta_u|G_u=g,\alpha_g) \\
   &\propto \prod_{t=1}^T \theta_{ut}^{Z_{-d,t}^{(u)}} \prod_{t=1}^T \theta_{ut}^{\alpha_{gt}-1} \\
   &= \prod_{t=1}^T \theta_{ut}^{Z_{-d,t}^{(u)} + \alpha_{gt} - 1} \\
   &\propto \text{Dir}(\vect{Z_{-d}^{(u)}} + \alpha_g) 
\end{align*}

Using this result in the first integral, we have: 

\begin{align*}
    \int \theta_{ut}p(\theta_u|\vect{Z_{-ud}})\ d\theta_u &= \frac{\Gamma \left(\sum_j Z_{-d,j}^{(u)} + \alpha_{gj}\right)}{\prod_j \Gamma\left(Z_{-d,j}^{(u)} + \alpha_{gj}\right)} \int \theta_{ut} \prod_{j=1}^T \theta_{uj}^{Z_{-d,j}^{(u)} + \alpha_{gj} - 1} \ d\theta_u \\
    &= \frac{\Gamma \left(\sum_j Z_{-d,j}^{(u)} + \alpha_{gj}\right)}{\prod_j \Gamma\left(Z_{-d,j}^{(u)} + \alpha_{gj}\right)} \frac{\prod_j \Gamma\left(Z_{-d,j}^{(u)} + \alpha_{gj} + I(j=t)\right)}{\Gamma \left(\sum_t Z_{-d,j}^{(u)} + \alpha_{gj} + 1\right)} \\
    &= \frac{Z_{-d,t}^{(u)} + \alpha_{gt}}{\sum_j Z_{-d,j}^{(u)} + \alpha_{gj}}
\end{align*}

This is the result in the paper. $\Gamma(\cdot)$ denotes the gamma function and $I(\cdot)$ the indicator function. The integral is computed by recognizing the kernel of a Dirichlet distribution with the same parameters as the posterior with 1 added the $t$'th parameter.  

The second integral is computed similarly. For documents with $Z_{ud} = t$, the word counts are a multinomial draw of size $n$ from the probability distribution $\beta_t$. Thus, the posterior can be expressed as: $p(\beta_t|\vect{Z_{-ud}},\vect{W_{-ud}}) \propto p(\beta_t|\vect{W_{-ud}^{(t)}})$, conditioning only on the word counts from documents with $Z_{ud}=t$. With a $\text{Dir}(\eta)$ prior on $\beta_t$, the posterior also Dirichlet with parameters $\vect{W_{-ud}^{(t)}} + \eta$ by the same conjugate update shown above. Thus, the second integral simplifies to:

\begin{align*}
    \int \prod_i \beta_{ti}^{W_{ud,i}} p(\beta_t|\vect{Z_{-ud}},\vect{W_{-ud}}) \ d\beta_t &= \frac{\Gamma\left(N\eta + \sum_i^N W_{-ud,i}^{(t)}\right)}{\prod_i^N \Gamma\left(\eta + W_{-ud,i}^{(t)}\right)} \int \prod_i \beta_{ti}^{W_{ud,i}} \prod_i \beta_{ti}^{W_{-ud,i} + \eta} \ d\beta_t \\
    &= \frac{\Gamma\left(N\eta + \sum_i^N W_{-ud,i}^{(t)}\right)}{\prod_i^N \Gamma\left(\eta + W_{-ud,i}^{(t)}\right)} \frac{\prod_i^N \Gamma\left(\eta + W_{-ud,i}^{(t)} + W_{ud,i}\right)}{\Gamma\left(N\eta + \sum_i^N W_{-ud,i}^{(t)} + W_{ud,i}\right)}
\end{align*}

This is the result in the paper. The integral is computed by recognizing the kernel of a Dirichlet distribution with the same parameters as the posterior for $\beta_t$ with the word counts for tweet $ud$ added to the respective parameters. Some simplification of the gamma functions are possible. They do not add to the intuitive understanding, so we do not show them here.

\section{Variational stLDA-C}

This section derives the variational approximation to the posterior from Section 3.2 in the paper. This follows the same general procedure as the variational approximation derived in Blei et al. (2003), and we note the similarities where appropriate. 

\subsection{ELBO Bound}

Here we derive a evidence lower bound (ELBO) on the KL divergence between the variational distribution and the posterior.  First, we will build the ELBO for an individual user after re-introducing the notation from the main paper. 

$\vect W_u$ is the ($n_u \times V$) document-term matrix for user $u$. The latent variables are: $g_u$ their cluster membership, $\theta_u$ their topic distribution, and $\vect Z_u$ the topics of their posts. For simplicity, I omit the $\_u$ notation for the remainder of this section. The model parameters are: $\alpha$ the $G\times T$ matrix of Dirichlet parameters for each cluster, $\beta$ the $T \times V$ matrix of topic distributions, and $\xi$ the corpus-level cluster proportions. The variational distribution $q$ is expressed as: 

\[q(g,\theta,\vect Z|\lambda,\gamma,\phi) = q(g|\lambda)q(\theta|\gamma)\prod_{d=1}^{n_u} q(z_d|\phi_d)\]

Here $q(g|\lambda)$ is Cat($\lambda$), $q(\theta|\gamma)$ is Dir($\gamma$), and $q(z_d|\phi_d)$ is Cat($\phi_d$). Both $\lambda$ and $\phi_d$ are points on the probability simplex; $\gamma$ is a $T$ dimensional vector of positive numbers. Although not made explicit in the notation, $\{\lambda,\gamma,\phi\}$ are all \textit{user-specific} quantities. For consistency with Blei et al. (2003) notation, I let $\xi$ denote the cluster mixing proportions ($\phi$ in the main paper) and let $\phi_d$ be the post-specific variational parameters. Now, we express the probability of the user's posts. 

\begin{align*}
    \log p(\vect W| \alpha,\beta,\xi) &= \log \sum_{g} \int_{\theta} \sum_z p(\theta,\vect Z,g|\alpha,\beta,\xi) \\
    &= \log \sum_{g} \int_{\theta} \sum_z p(\theta,\vect Z,g|\alpha,\beta,\xi) \frac{q(g,\theta,\vect Z)}{q(g,\theta,\vect Z)} \\
    &= \log E_q\left[\frac{p(\theta,\vect Z,g|\alpha,\beta,\xi)}{q(g,\theta,\vect Z)}\right] \\
    &\geq E_q[\log p(g,\theta,\vect Z,\vect W|\alpha,\beta,\xi)] -E_q[\log q(g,\theta,\vect Z)] := L
\end{align*}

The last step is from applying Jensen's inequality to push the log through the expectation. $E_q$ notes the expectation is taken with respect to the variational distribution $q$. 

As in normal LDA, the difference between the left and right hand terms ($\log p(\vect W| \alpha,\beta,\xi)$ and the lower bound, $L$) is the KL divergence between the variational posterior and the true posterior. Let $D(A||B)$ denote the KL divergence between A and B. 

\begin{align*}
    E_q[\log p(g,\theta,\vect Z,\vect W|\alpha,\beta,\xi)] -E_q[\log q(g,\theta,\vect Z)] &= \log p(\vect W|\alpha,\beta,\xi) + E_q[\log p(g,\theta,\vect Z|\vect W,\alpha, \beta,\xi) - E_q[\log q(g,\theta,\vect Z)] \\
    &= \log p(\vect W|\alpha,\beta,\xi) - D(q(g,\theta,\vect Z)||p(g,\theta,\vect Z|\vect W,\alpha, \beta,\xi))
\end{align*}

The likelihood of the data and latent parameters can be factored into simpler conditional distributions. 

\[ p(g,\theta,\vect Z,\vect W|\alpha,\beta,\xi) = p(g|\xi)p(\theta|g,\alpha)p(\vect Z | \theta)p(\vect W|\vect Z,\beta)\] 

Thus, the ELBO (L) can be expressed as (reintroducing variational parameters): 

\begin{align*}
    L(\lambda,\gamma,\phi;\alpha,\beta,\xi) &= E_q[\log p(g,\theta,\vect Z,\vect W|\alpha,\beta,\xi)] -E_q[\log q(g,\theta,\vect Z)] \\
    &= E_q[\log p(g|\xi)] + E_q[\log p(\theta|g,\alpha)] + E_q[\log p(\vect Z|\theta)] + E_q[\log p(\vect W|\vect Z,\beta)] \\
    &- E_q[\log q(g)] - E_q[\log q(\theta)] - E_q[\log q(\vect Z)]
\end{align*}

Each expectation can be simplified. Each line below corresponds to expansion of the seven terms above, in order. %Note that for $\theta\sim Dir(\gamma)$, $E[\log \theta_i] = \Psi(\gamma_i)-\Phi(\sum_i \gamma_i)$ where $\Phi$ is XXX. 
\begin{align}
    L(\lambda,\gamma,\phi;\alpha,\beta,\xi) &= \sum_g \lambda_g \log(\xi_g) \label{eq:elboU_start}\\
    &+ \sum_g \lambda_g \left(\log \Gamma(\sum_{t=1}^T \alpha_{gt}) - \sum_{t=1}^T \log \Gamma(\alpha_{gt}) + \sum_{t=1}^T (\alpha_{gt}-1) E_q[\log \theta_t]\right)  \label{eq:elboU_alpha}\\
    &+ \sum_{d=1}^{n} \sum_{t=1}^T \phi_{dt} E_q[\log\theta_t] \\
    &+ \sum_{d=1}^n \sum_{t=1}^T   \phi_{dt}\left[\log\left(\frac{\left(\sum_{j=1}^V W_{dj}\right)!}{\prod_{j=1}^V W_{dj}!}\right) + \sum_{j=1}^V W_{dj} \log \beta_{tj}\right] \\
    &- \sum_g \lambda_g \log \lambda_g \\
    &- \log \Gamma(\sum_{t=1}^T \gamma_t) + \sum_{t=1}^T \log \Gamma(\gamma_t) - \sum_{t=1}^T(\gamma_t -1) E_q[\log \theta_t] \\
    &- \sum_{d=1}^n\sum_{t=1}^T \phi_{dt} \log(\phi_{dt}) \label{eq:elboU_end}
\end{align}

These are nearly the same terms derived by Blei et al. (2003) Appendix A.3, Equation (15). Lines (3), (6), and (7) are the same. Lines (1) and (5) are new for cluster membership. (2) is more complicated because of the summation over the clusters; the $\sum_g$ component is new. (4) is only different because the normalizing constant for the multinomial distribution is added; the post is a draw of size $n_d$ from the relevant topic distribution, rather than each word being a draw of size one. Line (4) is the only change required if stLDA (no clustering) is used rather than traditional LDA. 

\subsection{Variational Parameter Estimation ($\lambda,\gamma,\phi$)}

Here I derive the ELBO-maximizing values of each variational parameter. Note that, as with vanilla LDA, the update rules are dependent, so the estimation needs to iterate between all of them until convergence. This section relies on computing $E_q[\log \theta_t]$. The value is expressed algebraically in Blei et al. (2003) Appendix A.1. 

\subsubsection{$\lambda$}

Recall that $\lambda_g$ is the variational probability that the user is in cluster $g$. Maximizing $L$ with respect to $\lambda$ only involves lines (1), (2), and (5) with the constraint that $\sum_g \lambda_g =1$. The derivative of $L$ with respect to $\lambda_g$ is: 

\[\log(\xi_g) + f(\alpha_g,\gamma) - \log \lambda_g - 1 + c\]

Where $f(\alpha_g,\gamma)$ is from line (2) above and $c$ is a constant from the Lagrangian. Setting equation equal to zero gives 

\[\lambda_g \propto \xi_g \exp(f(\alpha_g,\gamma)) = \xi_g \exp(E_q[\log p(\theta|g,\alpha_g)])\]

This is interpretable as proportional to the ``prior'' $\xi_g$ times the exponential of the variational expected log likelihood of $\theta$ if $\theta\sim Dir(\alpha_g)$. This is interpretable as computing $p(g|\theta)$ as $\propto p(\theta|g)p(g)$ with the likelihood approximated via the variational posterior. 

\subsubsection{$\phi$}

Recall that $\phi_{dt}$ is the variational probability that document $d$ has topic $t$. Maximizing $L$ with respect to $\phi$ only involves lines (3), (4), and (7) with the constraint that $\forall d$, $\sum_t \phi_{dt} = 1$. The terms in $L$ that are functions of $\phi_{dt}$ are isolated below, with $c_d$ to represent the multinomial normalizing constant. 

\[\phi_{dt} E_q[\log\theta_t] + \phi_{dt} \log(c_d) +  \sum_{j=1}^V W_{dj}\log \beta_{tj} - \phi_{dt} \log(\phi_{dt})\]

The derivative of $L$ with respect to $\phi_{dt}$ including Lagragian term $\zeta$ is: 

\[E_q[\log \theta_t] + \log(c_d) + \sum_{j=1}^V W_{dj}\log\beta_{tj} - \log(\phi_{dt}) -1 + \zeta \]

Setting equal to zero gives: 

\[\log \phi_{dt} = E_q[\log \theta_t] + \log(c_d) + \log \left(\prod_{j=1}^V \beta_{tj}^{W_{dj}}\right) -1 + \zeta \implies \]
\[\phi_{dt} \propto \exp(E_q[\log \theta_t]) \prod_{j=1}^V \beta_{tj}^{W_{dj}}\]

This is again interpretable as the likelihood times the prior, but this time the prior is approximated via the variational distribution.

\subsubsection{$\gamma$}

Maximizing $L$ with respect to $\gamma$ involves lines (2), (3), and (6), noting that $E_q[\log \theta_t] = g_t(\gamma)$, a known formula with well established numerical approximation methods. The components of $L$ involving $\gamma_t$ are listed below:

\[\sum_g \lambda_g \sum_t (\alpha_{gt}-1)g_t(\lambda) + \sum_d \sum_t \phi_{dt}g_t(\lambda) - \log \Gamma\left(\sum_{t=1}^T \gamma_t\right) + \log \Gamma(\gamma_t) - \sum_{t=1}^T(\gamma_t -1) g_t(\lambda)\]

This simplifies to:\footnote{The -1 parenthetical terms cancel because $\sum_g \lambda_g = 1$} 

\[\sum_t g_t(\lambda)\left(\sum_g \lambda_g \alpha_{gt} + \sum_d \phi_{dt} -\gamma_t\right) - \log \Gamma \left(\sum_t \gamma_t\right) + \log \Gamma(\gamma_t)\]

Before taking the derivative, it is necessary to expand $g_t$. $g_t(x) = \Psi(x_t) - \Psi(\sum_t x_t)$ where $\Psi$ is the digamma function (first derivative of the log gamma function). Note that this is the same expression as in Blei et al. (2003) A.3.2 with $\sum_g \lambda_g \alpha_{gt} := \alpha_i$, the sum of each cluster's relevant Dirichlet parameter weighted by (variational) probability of cluster membership $\lambda_g$. This will result in the same maximizing value with the substitution because the substitution just changes a constant term. For clarity, however, I will continue to fully derive the result here. For ease of notation, let $\widetilde{\alpha}_t = \sum_g \lambda_g \alpha_{gt}$. 

Substituting the expression for $g_t$ and $\widetilde{\alpha}_t$ gives: 

\[\sum_t \left(\widetilde{\alpha}_t + \sum_d \phi_{dt} -\gamma_t\right)\left(\Psi(\lambda_t)-\Psi\left(\sum_t \gamma_t\right)\right) - \log \Gamma \left(\sum_t \gamma_t\right) + \log \Gamma(\gamma_t)\]

Taking a derivative with respect to $\gamma_t$ gives:

\begin{align*}
    \frac{\partial L }{\partial \gamma_t} =& \left(\widetilde{\alpha}_t + \sum_d \phi_{dt} - \gamma_t\right)\Psi'(\gamma_t) -\Psi(\gamma_t) \\
    & -\sum_i \left(\widetilde{\alpha}_i + \sum_d \phi_{di} - \gamma_i\right)\Psi'\left(\sum_i \gamma_i\right) + \Psi'\left(\sum_i \gamma_i\right)\\
    &-\Psi\left(\sum_t \gamma_t\right) + \Psi (\gamma_t) \\
    =& \left(\widetilde{\alpha}_t + \sum_d \phi_{dt} - \gamma_t\right)\Psi'(\gamma_t) -\sum_i \left(\widetilde{\alpha}_i + \sum_d \phi_{di} - \gamma_i\right)\Psi'\left(\sum_i \gamma_i\right)
\end{align*}

Setting equal to zero yields a maximum (for all $t$) of $\gamma_t = \sum_g \lambda_g \alpha_{gt} + \sum_d \phi_{dt}$.

\subsection{Model Parameter Estimation ($\alpha,\beta,\xi$)}

The above section was just for a single user (the analog of a single document in traditional LDA). To estimate the parameters shared across users, cluster-specific Dirichlet parameters $\alpha_g$, topic distributions $\beta$, and cluster proportions $\xi$, we have to sum the ELBO bound derived above across users. In another abuse of notation return the $\_u$ do denote individual users. 

Let the full likelihood be expressed as: $$L(\alpha,\beta,\xi) = \sum_u L_u(\lambda_u,\gamma_u,\phi_u;\alpha,\beta,\xi)$$

$L_u$ is the function defined by equations $(\ref{eq:elboU_start})-(\ref{eq:elboU_end})$ above with the user-dependence made explicit. This is the function that needs to be maximized with respect to $\{\alpha,\beta,\xi\}$ holding $\{\lambda_u,\gamma_u,\phi_u\}$ constant. 

\subsubsection{$\xi$}

This is the easiest parameter. It appears only in term (\ref{eq:elboU_start}). Taking a derivative and using a Lagrangian to enforce $\sum_g \xi_g = 1$ gives: 

\[\xi_g \propto \sum_u \lambda_{ug} \]

\subsubsection{$\beta$}

This is similar to the variational multinomial in A.4.1 of Blei et al. (2003), with only the added complexity of having multiple words drawn from the same topic. 

The full likelihood terms with $\beta$ plus Lagrangian term is expressed below. $d\in\mathcal{D}_u$ indexes over all documents produced by user $u$. %$c_{ud}$ is the multinomial normalizing constant for user $u$'s $d$th document.

\[\sum_u \sum_{d\in\mathcal{D}_u} \sum_t \sum_j  \phi_{udt} W_{udj} \log \beta_{tj} - \sum_t \zeta_t\left(1-\sum_j^V \beta_{tj}\right)\]

Taking derivatives with respect to $\beta_{tj}$ results in: 

\[\beta_{tj} \propto \sum_u \sum_{d\in\mathcal{D}_u} \phi_{udt}W_{udj}\]

The intuition is that $\beta_t$ is proportional to the word occurrences weighted by the variational probability that the word was drawn from topic $t$. 

\subsubsection{$\alpha$}

These terms are more complicated but can be computed in parallel over clusters. Recall that $\alpha$ is a $G\times T$ matrix of positive values. The ELBO terms involving $\alpha$ are in line $(\ref{eq:elboU_alpha})$. 

\[\sum_u \sum_g \lambda_{ug} \left(\log \Gamma(\sum_{t=1}^T \alpha_{gt}) - \sum_{t=1}^T \log \Gamma(\alpha_{gt}) + \sum_{t=1}^T (\alpha_{gt}-1) g_t(\gamma_{u})\right) \]

Let $\Psi$ denote the digamma function, the first derivative of the log gamma function. Taking a derivative with respect to $\alpha_{gt}$ gives: 

\[\frac{\partial L}{\partial \alpha_{gt}} = \sum_u \lambda_{ug} \left[\Psi\left(\sum_t \alpha_{gt} \right) - \Psi(\alpha_{gt}) + g_t(\gamma_u) \right] = M_g \left[\Psi\left(\sum_t \alpha_{gt} \right) - \Psi(\alpha_{gt})\right] + \sum_u \lambda_{ug} g_t(\gamma_u)\]

Where $M_g = \sum_u \lambda_{ug}$. Note that this depends on $\alpha_{gt'}$ for $t'\neq t$ but not on any $\alpha_{g't'}$ where $g'\neq g$. Thus, the update step can be computed simultaneously for each cluster. 

This has the same form as the derivative in traditional LDA so the same Newton-Raphson algorithm can be used based on the structure of the Hessian matrix. 

\[\frac{\partial L}{\partial \alpha_{gt}\alpha_{gt'}} = -I(t=t')M_g\Psi'(\alpha_{gt}) + M_g \Psi'\left(\sum_t \alpha_{gt}\right)\]

Thus, the Hessian for cluster $g$ can be expressed as: $\vect H_g = diag(-M_g \Psi'(\vect \alpha_g)) + \vect 1 \vect 1^T M_g\Psi'(\sum_t \alpha_{gt})$, a diagonal matrix plus a constant matrix. 

%{\color{blue} Note, this looks slightly different from the final derivation in Blei et al. (2003). I think they have a mistake in the final two lines in the last appendix. The result should be the above with $M$ replacing $M_g$.}